\documentclass[onecolumn,showpacs,preprintnumbers,elsart]{revtex4}
\usepackage{amsmath}
\usepackage{mathrsfs}
\usepackage{graphicx}
\usepackage{dcolumn}
\usepackage{bm}
\usepackage[center]{subfigure}
\usepackage{color}
\usepackage{ifpdf}

\begin{document}

\title{Wave modes trapped in rotating nonlinear potentials}
\author{Yongyao Li$^{1,2}$, Wei Pang$^{3}$, and Boris A. Malomed$^{1}$}
\affiliation{$^{1}$Department of Physical Electronics, School of Electrical Engineering,
Faculty of Engineering, Tel Aviv University, Tel Aviv 69978, Israel\\
$^{2}$Department of Applied Physics, South China Agricultural University,
Guangzhou 510642, China \\
$^{3}$ Department of Experiment Teaching, Guangdong University of
Technology, Guangzhou 510006, China.}

\begin{abstract}
We study modes trapped in a rotating ring with the local strength of the
nonlinearity modulated as $\cos \left( 2\theta \right) $, where $\theta $ is
the azimuthal angle. This modulation pattern may be of three different
types: self-focusing (SF), self-defocusing (SDF), and alternating SF-SDF.
The model, based on the nonlinear Schr\"{o}dinger (NLS) equation with
periodic boundary conditions, applies to the light propagation in a twisted
pipe waveguide, and to a Bose-Einstein condensate (BEC) loaded into a
toroidal trap, under the action of the rotating nonlinear \textit{%
pseudopotential} induced by means of the Feshbach resonance in an
inhomogeneous external field. In the SF, SDF, and alternating
regimes, four, three, and five different types of stable trapped
modes are identified, respectively: even, odd, second-harmonic (2H),
symmetry-breaking, and 2H-breaking ones. The shapes and stability of
these modes, together with transitions between them, are
investigated in the first rotational Brillouin zone. Ground-state
modes are identified in each regime. Boundaries between symmetric
and asymmetric modes are also obtained in an analytical form, by of
a two-mode approximation.
\end{abstract}

\maketitle

%\email{yongyaoli@gmail.com}
%\email{stszjy@mail.sysu.edu.cn}

%\preprint{APS/123-QED}

% Force line breaks with \\

\section{Introduction}

Optical and matter waves exhibit a plenty of dynamical scenarios under the
action of effective nonlinear potentials (which may sometimes be combined
with usual linear potentials) \cite{YVKRevMod}. The dynamics of such systems
is governed by the nonlinear Schr\"{o}dinger equation (NLSE) in optical
media, or Gross-Pitaevskii equation (GPE) in the context of Bose-Einstein
condensates (BECs). In either case, the nonlinear \emph{pseudopotential}
\cite{Harrison} may be induced by a regular \cite{LCQian,2D} or singular
\cite{singular}\ spatial modulation of the local nonlinearity. These systems
have been studied theoretically in a variety of one- \cite{LCQian,singular}
and two-dimensional (1D and 2D) \cite{2D} settings, and recently reviewed in
Ref. \cite{YVKRevMod}. To the same general class belong models which predict
that, in any dimension $D$, stable fundamental and vortex solitons can be
supported by a purely self-defocusing (SDF) nonlinearity growing towards the
periphery ($r\rightarrow \infty $) at any rate faster than $r^{D}$ \cite%
{Barcelona}.

In optics, such nonlinear potentials may be designed using the mismatch
between the nonlinearity of the host material and solid \cite{Fluan} or
liquid \cite{TTLarsen} stuff filling voids of photonic-crystal-fiber
waveguides. Another possibility to create the effective nonlinear potential
in optics is offered by inhomogeneous distributions of dopants which induce
the resonantly enhanced nonlinearity \cite{Kip}. In particular, it is
possible to use the Rhodamine B dopant added to the SU-8 polymer (a commonly
used transparent negative photoresist) \cite{Liangbing}, or Pr$^{3+}$ ions
doping the Y$_{2}$SiO$_{5}$ host medium \cite{Yongyao,Fleischhauer}.

In BEC, the pseudopotential can be created with the help of the Feshbach
resonance controlled by nonuniform magnetic \cite{SInouye} or optical \cite%
{optical} fields. In particular, the necessary pattern of the spatial
modulation of the scattering length, which determines the local strength of
the cubic nonlinearity in the respective GPE, can be induced by
appropriately designed magnetic lattices \cite{magnetic}.

Another well-known tool for the creation of various dynamical states
is provided by rotating potentials, which may trap optical and
matter waves. Effects of the rotation have drawn a great deal of
attention in the studies of BEC. A well-known results is that
rotational stirring of the condensate with repulsive interactions
leads to the formation of vortex lattices, see review
\cite{vort-latt}. Under special conditions (the compensation of the
trapping by the centrifugal force), giant vortices can be produced
too, which were studied in detail experimentally \cite{Cornell} and
theoretically \cite{giant}. In a binary immiscible BEC, vortex
streets were predicted to form, instead of the vortex lattices
\cite{street}. On the other hand, it was predicted that the rotation
of self-attractive condensates gives rise to several species of
localized modes, such as solitary vortices, mixed-vorticity states
(``crescents"), and quasi-solitons \cite{rotating-trap}.

It is natural too to consider the dynamics of matter-wave modes trapped in
rotating lattices, which can be created in BEC by broad laser beams
transmitted through a revolving mask \cite{sieve}. Quantum states and vortex
lattices have been studied in this setting \cite{in-sieve}, as well as the
depinning of trapped solitons and solitary vortices when the rotation rate
exceeds a critical value \cite{HS}. The nucleation of vortices in the
rotating lattice was demonstrated experimentally \cite{nucleation}.

In optics, a setting similar to the rotating lattice can be realized in
twisted photonic-crystal fibers \cite{twistedPC,Russell}. In plain optical
fibers, the twist affects the polarization dynamics \cite{RYChiao} and
couples it to the transmission of temporal solitons \cite{rocking}. In
helical photonic-crystal fibers, modified Bragg reflection and enhancement
of the mode conversion and transport have been studied \cite{Sujia}, and the
transformation of the linear moment of photons into the orbital angular
momentum has been demonstrated recently \cite{Russell}.

The simplest version of the rotating lattice is represented by the revolving
quasi-1D double-well potential (DWP). It gives rise to azimuthal Bloch bands
\cite{Ueda}, and allows one to support solitons and solitary vortices even
in the case of the SDF nonlinearity \cite{guiding}. The generation of a
vortex lattice in the rotating DWP was studied too \cite{WenLuo}. Further,
it is well known that the interplay of the DWP with the SF or SDF
nonlinearity provides for the simplest setting for the study of the
spontaneous symmetry breaking of even and odd states, respectively, in one
dimension \cite{SSB}. In this connection, a natural problem, which was
recently considered in Ref. \cite{we}, is a modification of the
symmetry-breaking scenarios in a rotating ring carrying the DWP potential,
along with the nonlinearity.

\begin{figure}[tbp]
\centering{\includegraphics[scale=0.6]{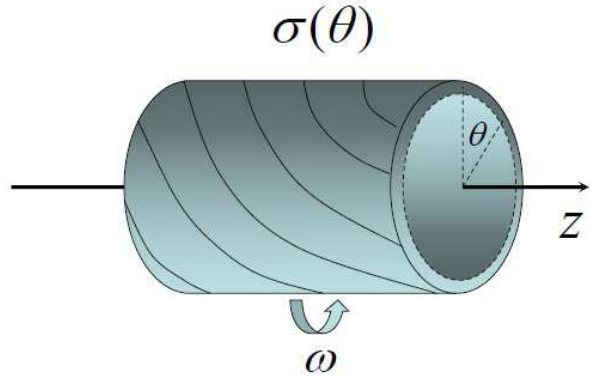}} \caption{(Color
online) The pipe nonlinear waveguide with the intrinsic
nonlinear potential, twisted at rate $\protect\omega $} \label{fig_1}
\end{figure}

While the dynamics of nonlinear waves trapped in rotating linear potentials
has been studied in detail, previous works did not tackle modes pulled by
rotating \textit{nonlinear} (pseudo-) potentials. This setting, which may be
implemented in optics and BEC alike, is the subject of the present work. To
analyze the basic features of the respective phenomenology, we here
concentrate on the 1D nonlinear potential on a rotating ring, as shown in
Fig. \ref{fig_1}. In optics, this system is realized as a pipe (hollow)
waveguide with an azimuthal modulation of the local nonlinearity, $\sigma
(\theta )$ ($\theta $ is the angular coordinate), \textit{twisted} with
pitch $2\pi /\omega $, where $\omega $ plays the role of the effective
rotation speed. In BEC, a similar setting corresponds to a toroidal trap,
which is available in the experiment \cite{torus}, combined with the
rotating nonlinear potential, that can be superimposed on top of the trap
\cite{Salasnich}. This combination realizes a \textit{rotating ring} \cite%
{Stringari} carrying the nonlinear potential.

It is necessary to define the sign of the nonlinear potential. In this work,
we consider three distinct cases, namely, the SF (attractive), SDF
(repulsive), and SF-SDF (alternating) nonlinearities, all subject to the
harmonic spatial modulation, which is defined as follows:
\begin{subequations}
\label{SF-SDF}
\begin{gather}
\sigma _{\mathrm{SF}}(\theta )=-\sin ^{2}\theta , \\
\sigma _{\mathrm{SDF}}(\theta )=\cos ^{2}\theta , \\
\sigma _{\mathrm{SF-SDF}}(\theta )=\cos \left( 2\theta \right) .
\end{gather}%
The solution domain is set as $-\pi \leq \theta \leq +\pi $. Note that, in
all the cases the local nonlinearity coefficient (\ref{SF-SDF}) has its
maxima at points $\theta =0$ and $\theta =\pm \pi $, and minima at $\theta
=\pm \pi /2$. We aim to find basic types of modes trapped in rotating
nonlinearity profiles (\ref{SF-SDF}), and establish their basic properties,
such as symmetry/asymmetry, stability, and the identification of the
respective ground states, varying two control parameters, \textit{viz}., the
rotation speed, $\omega $, and the total power (norm), $P$, which is defined
below in Eq. (\ref{P}).

The paper is structured as follows. In Section II, we formulate the system
and methods of the stability analysis of the modes. In Section III, we
present numerical results for the basic modes and their stability in each
type of the nonlinear potential (\ref{SF-SDF}). In Section IV, we present
analytical results, obtained by means of a two-mode approximation, which
explain a boundary between the symmetric and asymmetric modes in each case.
The paper is concluded by Section V.

\section{The model}

The dynamics of the optical wave (or the BEC wave function) in the rotating
ring is governed by the normalized one-dimensional NLSE (GPE), subject to
the periodic boundary conditions:
\end{subequations}
\begin{equation}
i{\frac{\partial }{\partial z}}\psi =\left[ -{\frac{1}{2}}{\frac{\partial
^{2}}{\partial \theta ^{2}}}+\sigma (\theta -\omega z)|\psi |^{2}\right]
\psi ,  \label{Eq5}
\end{equation}%
$\psi (\theta )\equiv \psi (\theta +2\pi )$, where $z$ is the propagation
distance in the case of the optical waveguide, and the radius of the ring is
scaled to be $1$. It is more convenient to rewrite Eq. (\ref{Eq5}) in the
rotating reference frame, with $\theta ^{\prime }\equiv \theta -\omega z$:
\begin{equation}
i{\frac{\partial }{\partial z}}\psi =\left[ -{\frac{1}{2}}{\frac{\partial
^{2}}{\partial \theta ^{\prime }{}^{2}}}+i\omega {\frac{\partial }{\partial
\theta ^{\prime }}}+\sigma (\theta ^{\prime })|\psi |^{2}\right] \psi ,
\label{Eq5p}
\end{equation}%
while the solution domain is defined as above, i.e., $-\pi \leq \theta
^{\prime }\leq +\pi $. For the BEC trapped in the rotating potential, the
respective GPE differs by replacing $z$ with time $t$.\ Equation (\ref{Eq5p}%
) conserves the total power (norm) of the field and its Hamiltonian
(energy),
\begin{equation}
P=\int_{-\pi }^{+\pi }\left\vert \psi (\theta ^{\prime })\right\vert
^{2}d\theta ^{\prime },  \label{Power}
\end{equation}%
\begin{equation}
H=\frac{1}{2}\int_{-\pi }^{+\pi }\left[ \left\vert {\frac{\partial \psi }{%
\partial \theta ^{\prime }}}\right\vert ^{2}+i\omega \left( \psi ^{\ast }{%
\frac{\partial \psi }{\partial \theta ^{\prime }}}-\psi {\frac{\partial \psi
^{\ast }}{\partial \theta ^{\prime }}}\right) +\sigma (\theta ^{\prime
})|\psi |^{4}\right] d\theta ^{\prime },  \label{Ham}
\end{equation}%
with the asterisk standing for the complex conjugate. Stationary modes with
real propagation constant $-\mu $ (in terms of BEC, $\mu $ is the chemical
potential) are sought for as
\begin{equation}
\psi \left( \theta ^{\prime },z\right) =\exp \left( -i\mu z\right) \phi
(\theta ^{\prime }),  \label{phi-psi}
\end{equation}%
with complex function $\phi \left( \theta ^{\prime }\right) $ obeying
equation
\begin{equation}
\mu \phi =\left[ -{\frac{1}{2}}{\frac{d^{2}}{d\theta ^{\prime }{}^{2}}}%
+i\omega {\frac{d}{d\theta ^{\prime }}}+\sigma (\theta ^{\prime })|\phi |^{2}%
\right] \phi .  \label{phi}
\end{equation}

The periodic boundary conditions, $\sigma (\theta ^{\prime }+2\pi )=\sigma
(\theta ^{\prime })$ and $\psi (\theta ^{\prime }+2\pi )=\psi (\theta
^{\prime })$, make Eq. (\ref{Eq5p}) invariant with respect to the \textit{%
boost transformation}, which allows one to change the rotation speed from $%
\omega $ to $\omega -N$ with arbitrary integer $N$:%
\begin{equation}
\psi \left( \theta ^{\prime },z;\omega -N\right) =\psi \left( \theta
^{\prime },z;\omega \right) \exp \left[ -iN\theta ^{\prime }+i\left( \frac{1%
}{2}{N}^{2}-N\omega \right) z\right] ,  \label{boost}
\end{equation}%
hence the speed may be restricted to interval $0\leq \omega <1$.
Furthermore, Eq. (\ref{Eq5p}) admits an additional invariance, relating
solutions with opposite signs of the speed: $\psi (\theta ^{\prime
},z;\omega )=\psi ^{\ast }(\theta ^{\prime },-z;-\omega )$. If combined with
boost $\omega \rightarrow \omega +1$, the latter transformation demonstrates
that the solutions with
\begin{equation}
\omega =1/2\pm \delta ,  \label{+-}
\end{equation}%
where $\delta <1/2$, are tantamount to each other. Thus, the rotation speed
may be eventually restricted to the fundamental interval,
\begin{equation}
0\leq \omega \leq 1/2,  \label{zone}
\end{equation}%
which plays the role of the first \textit{rotational Brillouin zone}, cf.
Ref. \cite{Ueda}.

The dynamical stability of the stationary solutions has been investigated
via numerical computation of eigenvalues for small perturbations, and
verified by direct simulations of the perturbed evolution. The perturbed
solutions are introduced as usual,
\begin{equation}
\psi =e^{-i\mu z}[\phi (\theta ^{\prime })+\varepsilon u(\theta ^{\prime
})e^{i\lambda z}+\varepsilon v^{\ast }(\theta ^{\prime })e^{-i\lambda ^{\ast
}z}],  \label{pert}
\end{equation}%
where $\varepsilon $ is an infinitesimal amplitude of the disturbance, $%
u(\theta ^{\prime })$ and $v(\theta ^{\prime })$ are the corresponding
eigenmodes, and $\lambda $ the eigenfrequency. The substitution of ansatz (%
\ref{pert}) into Eq. (\ref{Eq5p}) and linearization leads to the linear
eigenvalue problem,
\begin{equation}
\left(
\begin{array}{cc}
\mu {-}\hat{h}-i\omega \frac{\partial }{\partial \theta ^{\prime }} &
-\sigma \phi ^{2} \\
\sigma \left( \phi ^{\ast }\right) ^{2} & -\mu +\hat{h}-i\omega \frac{%
\partial }{\partial \theta ^{\prime }}%
\end{array}%
\right) \left(
\begin{array}{c}
u \\
v%
\end{array}%
\right) =\lambda \left(
\begin{array}{c}
u \\
v%
\end{array}%
\right) ,  \label{lambda}
\end{equation}%
where $\hat{h}=-(1/2)\partial ^{2}/\partial \left( \theta ^{\prime }\right)
^{2}+2\sigma |\phi |^{2}$ is the single-particle Hamiltonian. The underlying
solution $\phi $ is stable if Eq. (\ref{lambda}) generates solely real
eigenvalues. Lastly, stationary equation (\ref{phi}) was solved using
numerical code ``PCSOM" borrowed from Ref. \cite{YJK}.

\section{Numerical Results}

\subsection{The classification of trapped modes}

For all types of the nonlinear potentials defined in Eq. (\ref{SF-SDF}), the
numerical solution of stationary equation (\ref{phi}) with different inputs
(initial guesses) makes it possible to identify five basic species of
stationary states, which are listed in Table \ref{tab:guess}.

\begin{table}[tbp]
\caption{Different species of stationary modes, labeled by input waveforms
which generate them as solutions of Eq. (\protect\ref{phi}).}%
\begin{ruledtabular}
\begin{tabular}{lccr}
Inputs  & Types of modes   \\
\hline
 $\cos\theta'$ & Symmetric (even)\\
 $\sin\theta'$ & Antisymmetric (odd)\\
 $\sin^{2}\theta'$ & Second-harmonic (2H)\\
 $ b+\cos\theta',\,0<b\leq1$ & Symmetry-breaking\\
 $b+\sin\theta',\,0<b\leq1$ & 2H-breaking\\
\end{tabular}
\end{ruledtabular}
\label{tab:guess}
\end{table}

Following the symmetry of the nonlinearity-modulation patterns in Eq. (\ref%
{SF-SDF}), the modes are identified as symmetric (alias even) and
antisymmetric (alias odd), with respect to the central point,
$\theta ^{\prime }=0$. The mode of the ``second-harmonic" (2H, also
even) type refers to the dominant term in its Fourier decomposition.
The names of the last two modes in Table \ref{tab:guess} come from
the types of their symmetry breaking. In particular, the named of 2H-breaking mode is come from its spontaneous symmetry breaking via the 2H mode, which will be demonstrated in the following. Note that all the inputs
displayed in the table are real functions, while the numerically
found solutions of Eq. (\ref{phi}) with $\omega \neq 0$ are actually
complex. Naturally, real parts of the solutions generated by the
real inputs indicated in the table have the same parity (even/odd)
as the inputs, while the imaginary parts feature the opposite parity
(odd/even) (the input of the 2H-breaking type does not
feature a certain parity). It is shown below too that maxima and
minima of the local power (density) of the even mode defined in the
table coincide with the maxima and minima of the local nonlinearity,
$\sigma (\theta ^{\prime })$, while for the odd and 2H modes the
relation is the opposite, with the peak powers sitting in potential
wells [see Eq. (\ref{SF-SDF})], hence the odd and 2H modes tend to
have lower values of the energy, and may play the role of the ground
state, as confirmed below.

\subsection{The self-focusing nonlinearity}

\begin{figure}[tbp]
\centering%
\subfigure[] {\label{fig_2_a}
\includegraphics[scale=0.21]{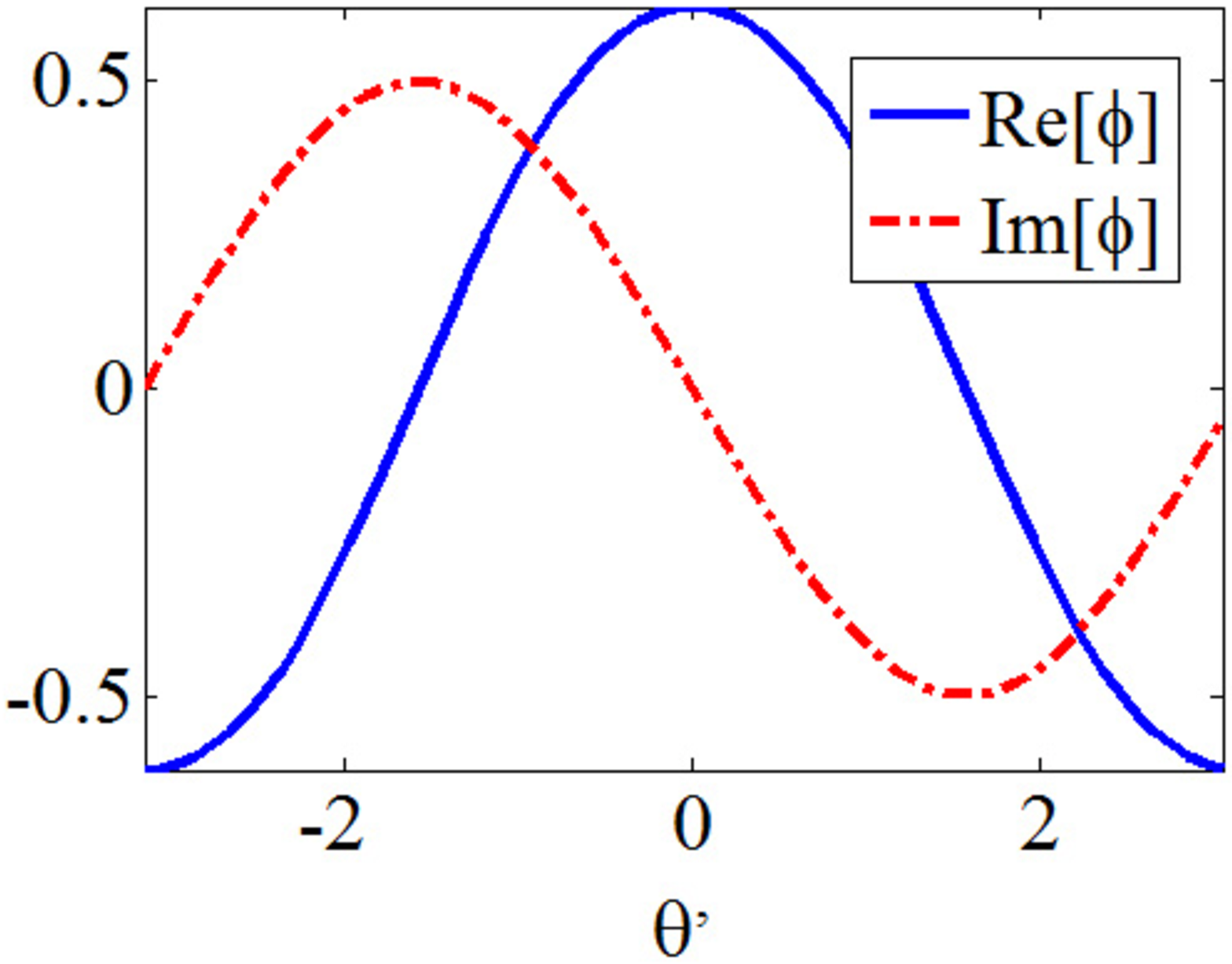}}%
\subfigure[] {\label{fig_2_b}
\includegraphics[scale=0.21]{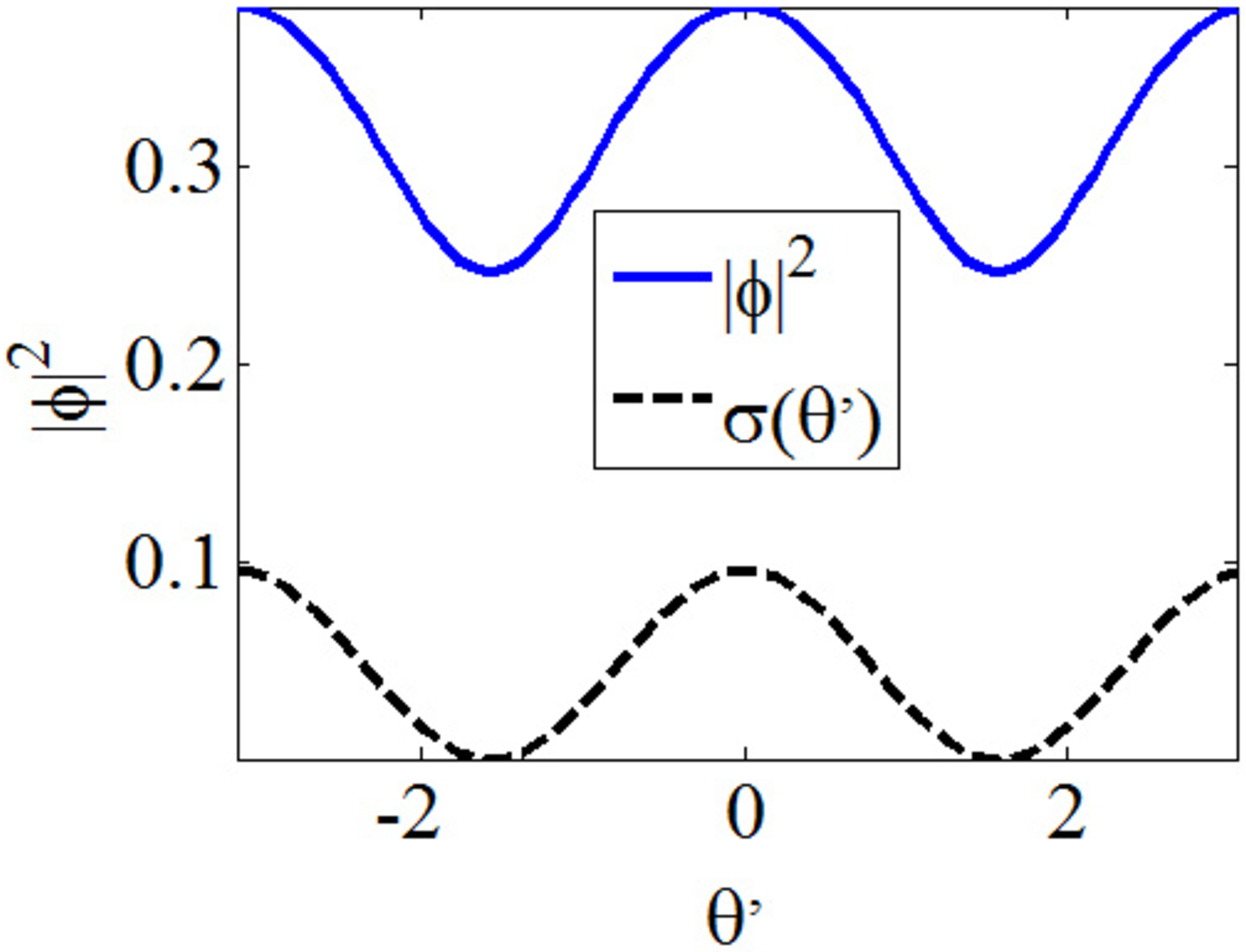}}
\subfigure[]{ \label{fig_2_c}
\includegraphics[scale=0.21]{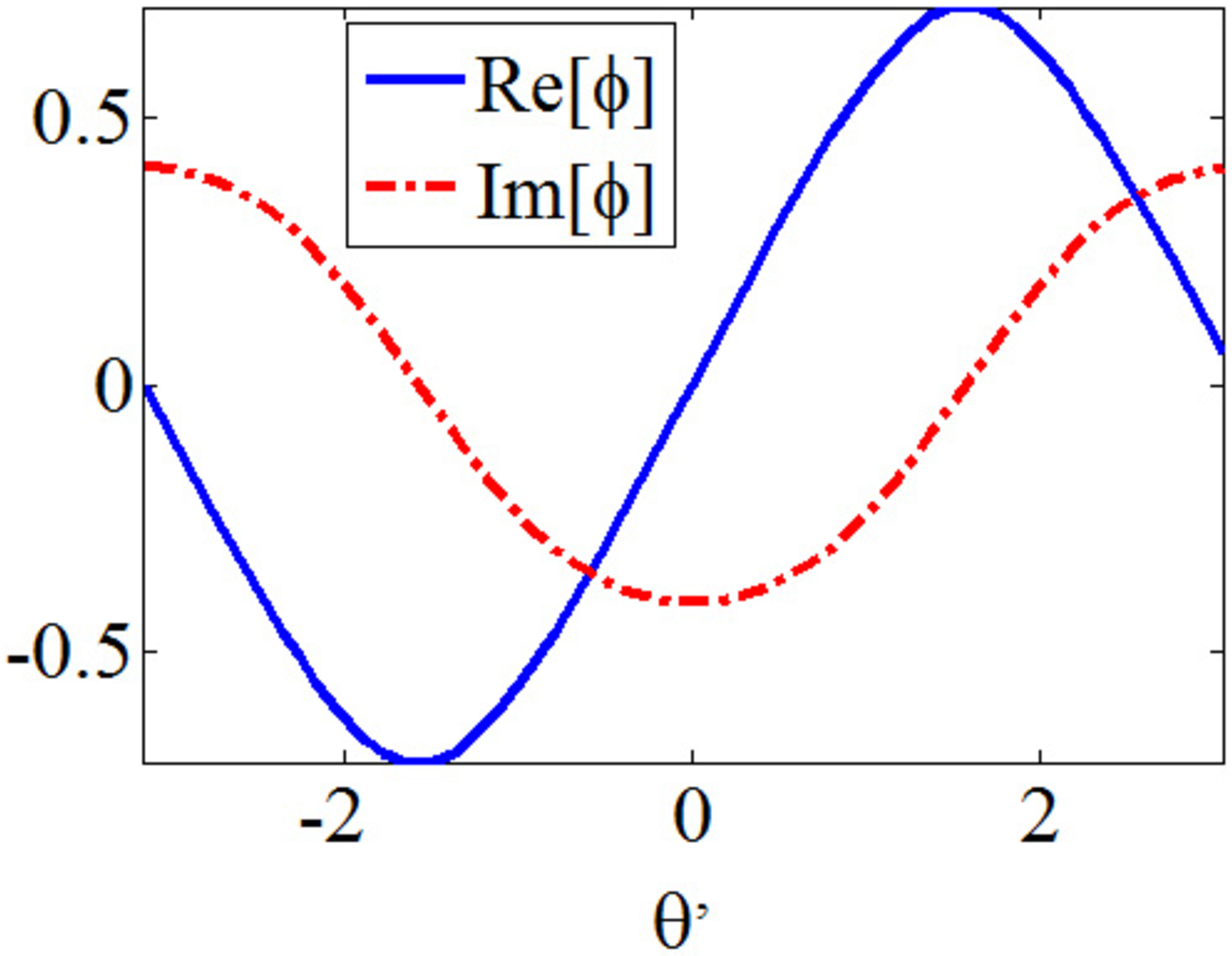}}%
\subfigure[]{ \label{fig_2_d}
\includegraphics[scale=0.21]{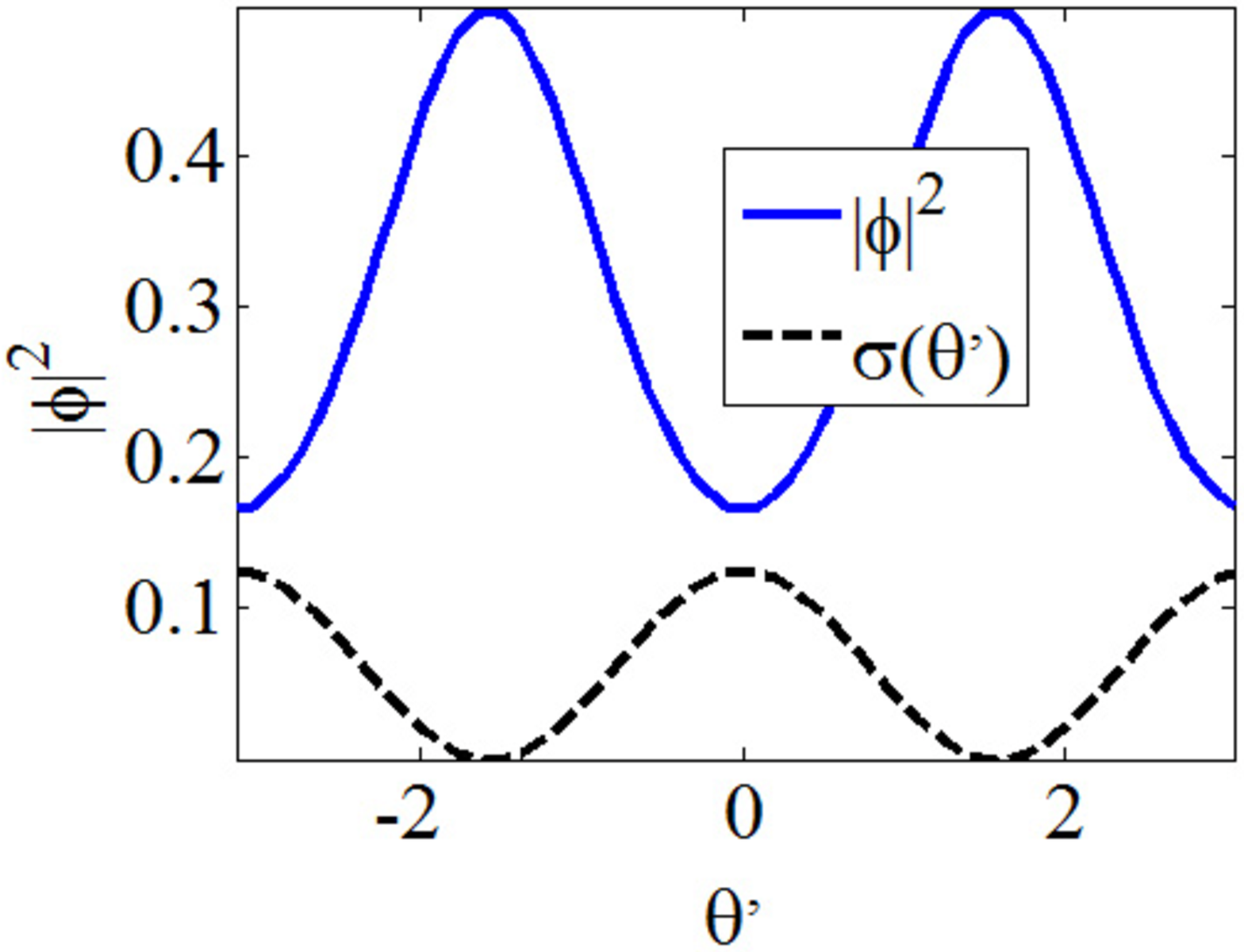}}
\caption{(Color online) {Examples of stable even and odd modes, found in the
system with self-focusing nonlinear potential (\protect\ref{SF-SDF}a), at
rotation speed $\protect\omega =0.25$, with total power $P_{\mathrm{even}}=P_{\mathrm{odd}}=2$. Panels (a),
(c) display, severally, real and imaginary parts of the even and odd modes,
while (b), (d) show their local-power (density) profiles. The dashed curves
in (b) and (d), and in similar panels displayed below, depict the
corresponding nonlinearity-modulation profile, $\protect\sigma (\protect%
\theta ^{\prime })$; in the present case, it is $\cos ^{2}\protect\theta %
^{\prime }$. }}
\label{SFIP2H}
\end{figure}

\begin{figure}[tbp]
\centering%
\subfigure[] {\label{fig_3_a}
\includegraphics[scale=0.21]{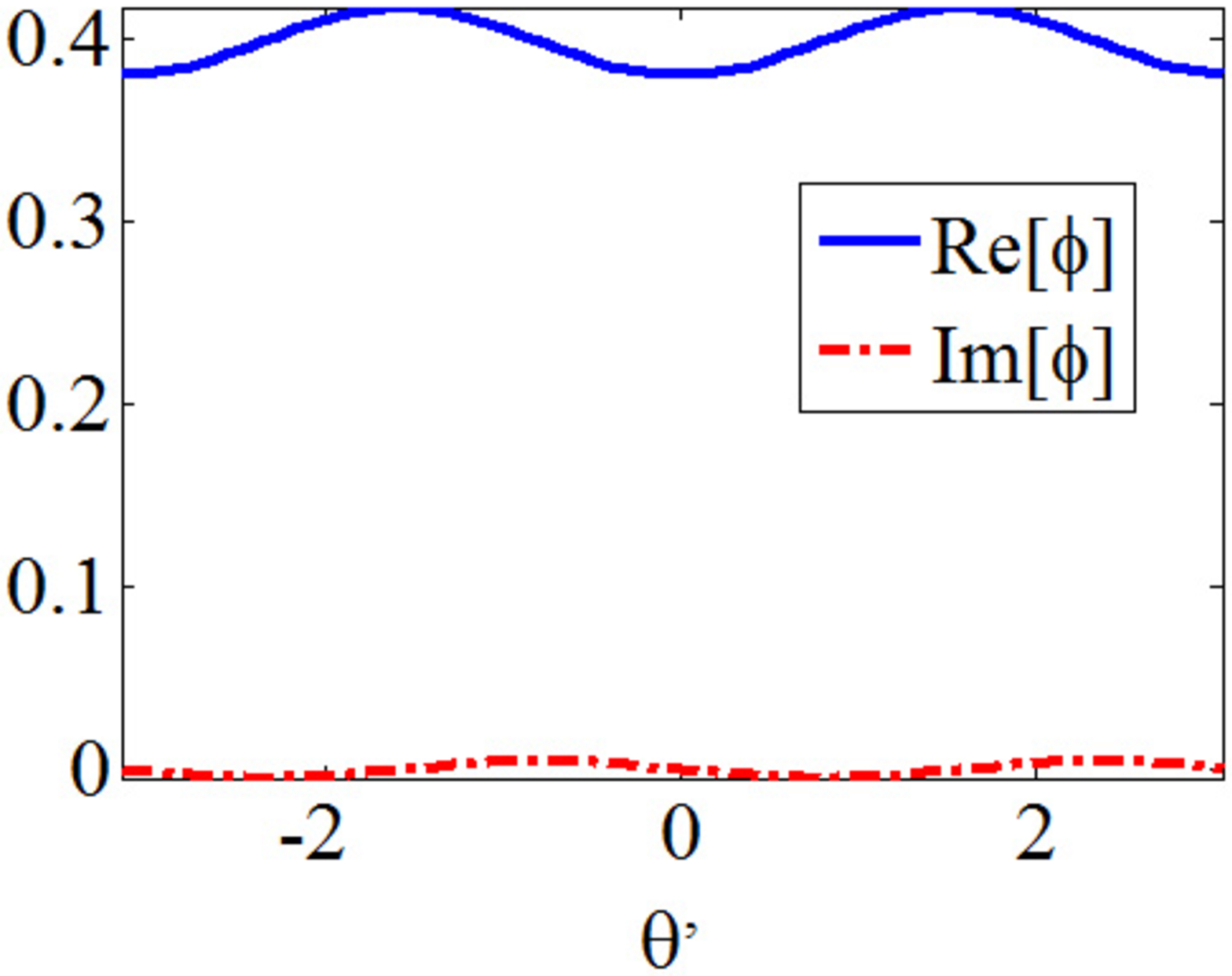}}%
\subfigure[] {\label{fig_3_b}
\includegraphics[scale=0.21]{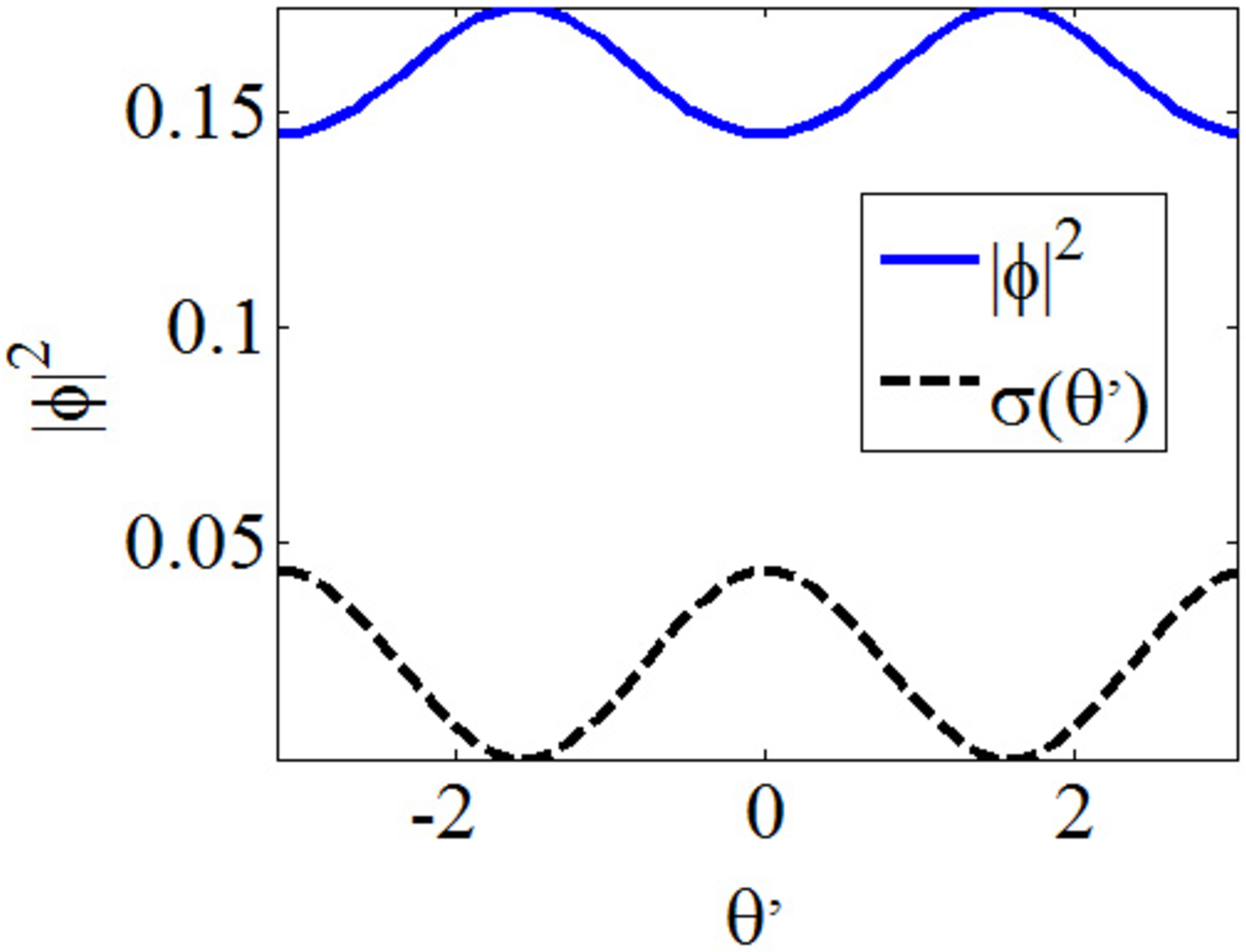}}
\subfigure[]{ \label{fig_3_c}
\includegraphics[scale=0.21]{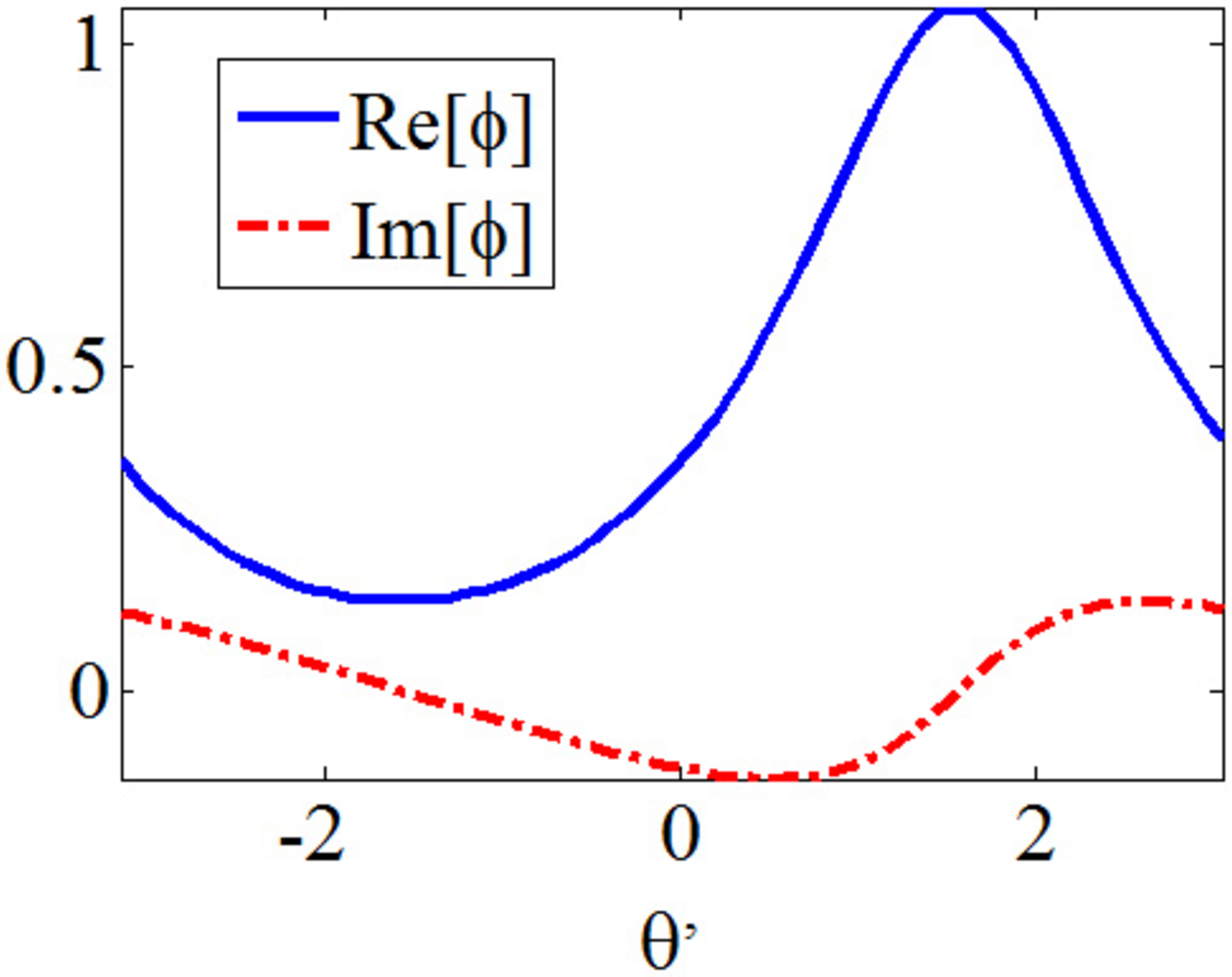}}
\subfigure[]{ \label{fig_3_d}
\includegraphics[scale=0.21]{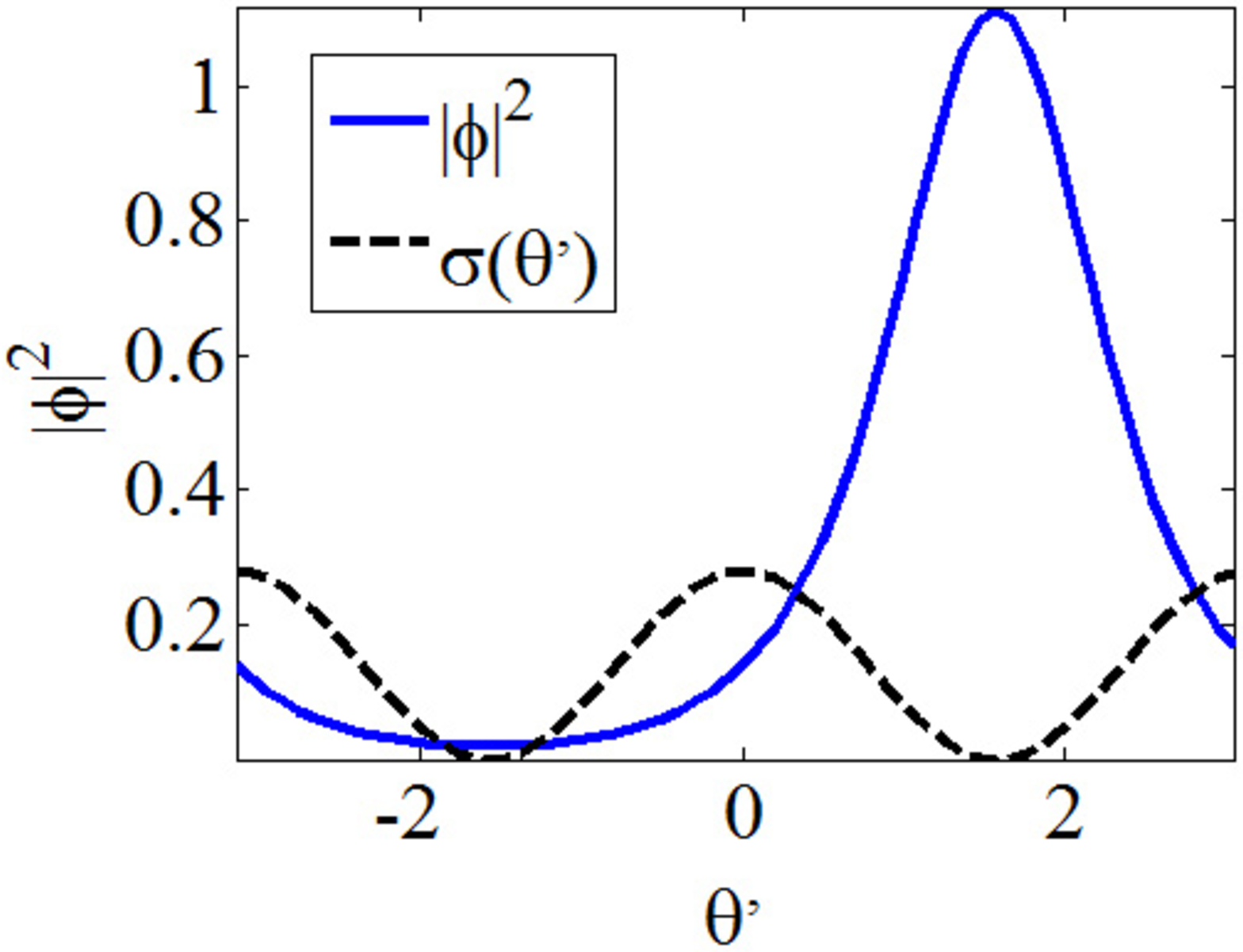}}
\caption{(Color online) {Examples of stable 2H and 2H-breaking modes, found
in the system with the self-focusing nonlinear potential at $\protect\omega %
=0.25$ and $P_{\mathrm{2H}}=1$ and $P_{\mathrm{2H-break.}}=2$, respectively. Panels have the same meaning as in
Fig. \protect\ref{SFIP2H}.}}
\label{SFOP}
\end{figure}

The SF nonlinear potential, defined as per Eq. (\ref{SF-SDF}a),
gives rise to four types of dynamically stable trapped modes,
\textit{viz}., even (symmetric), odd (antisymmetric), 2H and
2H-breaking ones (i.e., only the symmetry-breaking species is
missing in this case). Typical examples of these stable modes are
displayed in Figs. \ref{SFIP2H} and \ref{SFOP}. In addition to the
above-mentioned fact that the maxima and minima of the local power
coincide with those of the nonlinear potential for the even mode,
and, on the contrary, coincide with minima and maxima of the
potential for the odd and 2H modes, suggesting that either of the
latter modes may be a ground state, the figures demonstrate that the
2H-breaking mode has one maximum and one minimum of the local power,
both sitting in nonlinear-potential wells (minima).

\begin{figure}[tbp]
\centering%
\subfigure[] {\label{fig_4_a}
\includegraphics[scale=0.18]{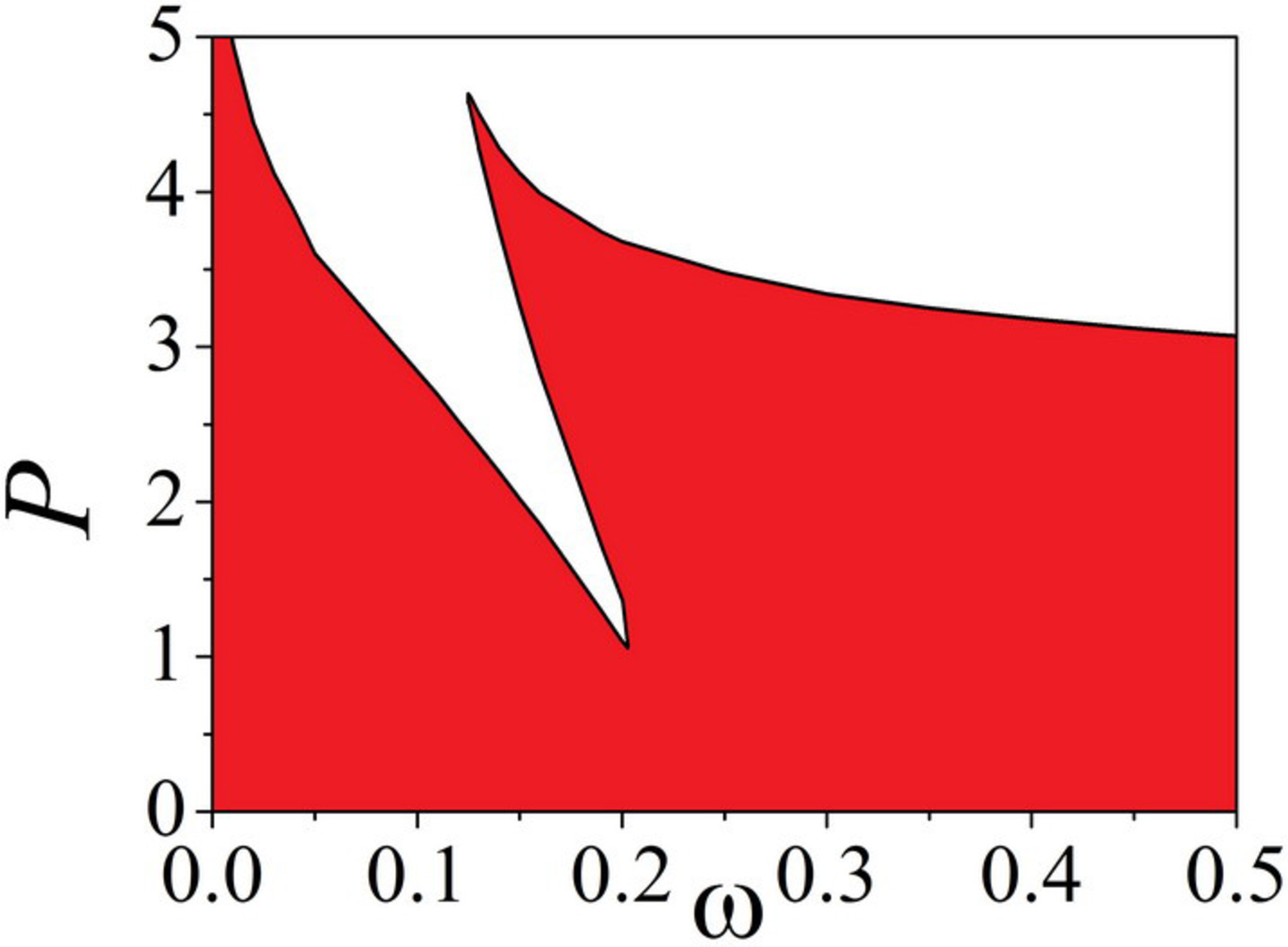}}%
\subfigure[] {\label{fig_4_b}
\includegraphics[scale=0.35]{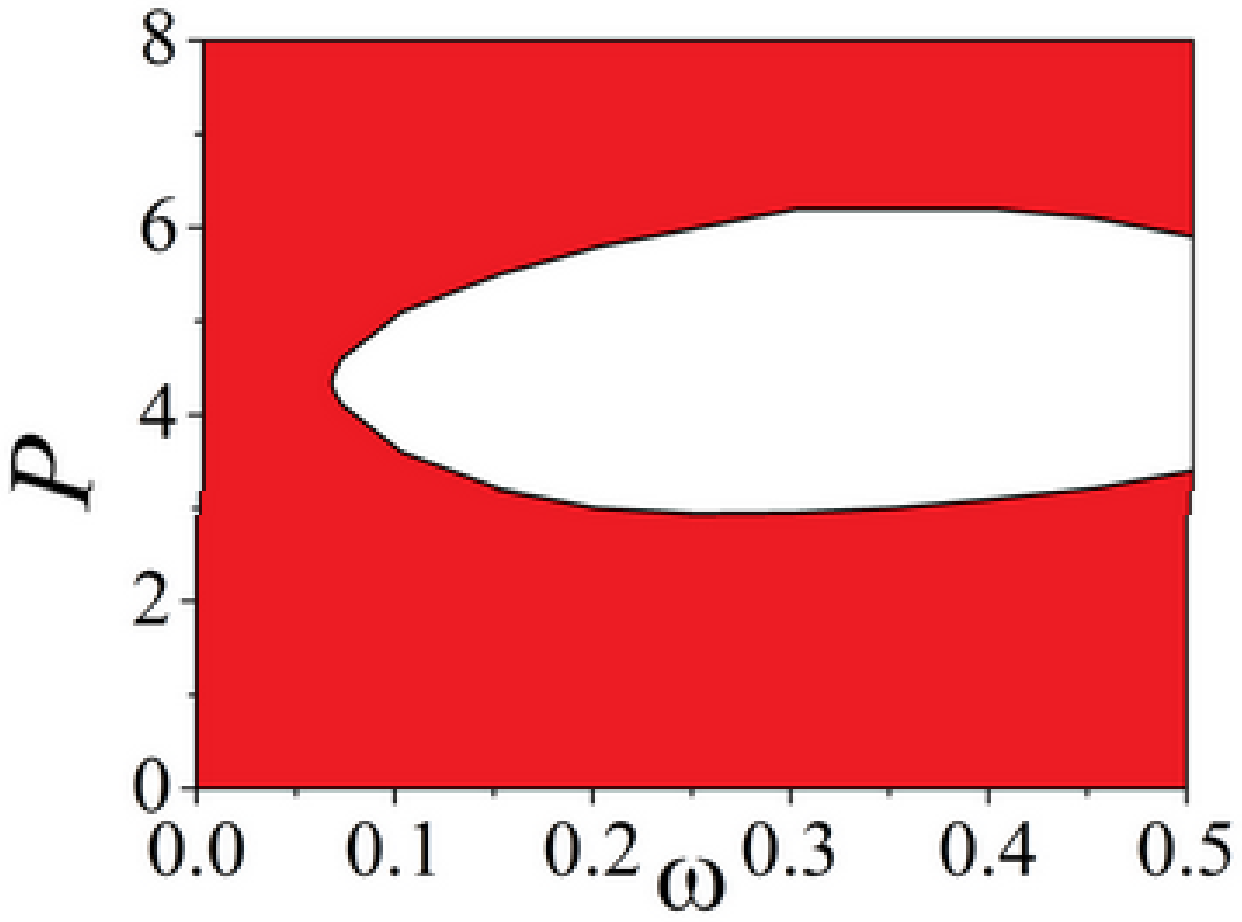}}
\subfigure[]{ \label{fig_35_c}
\includegraphics[scale=0.35]{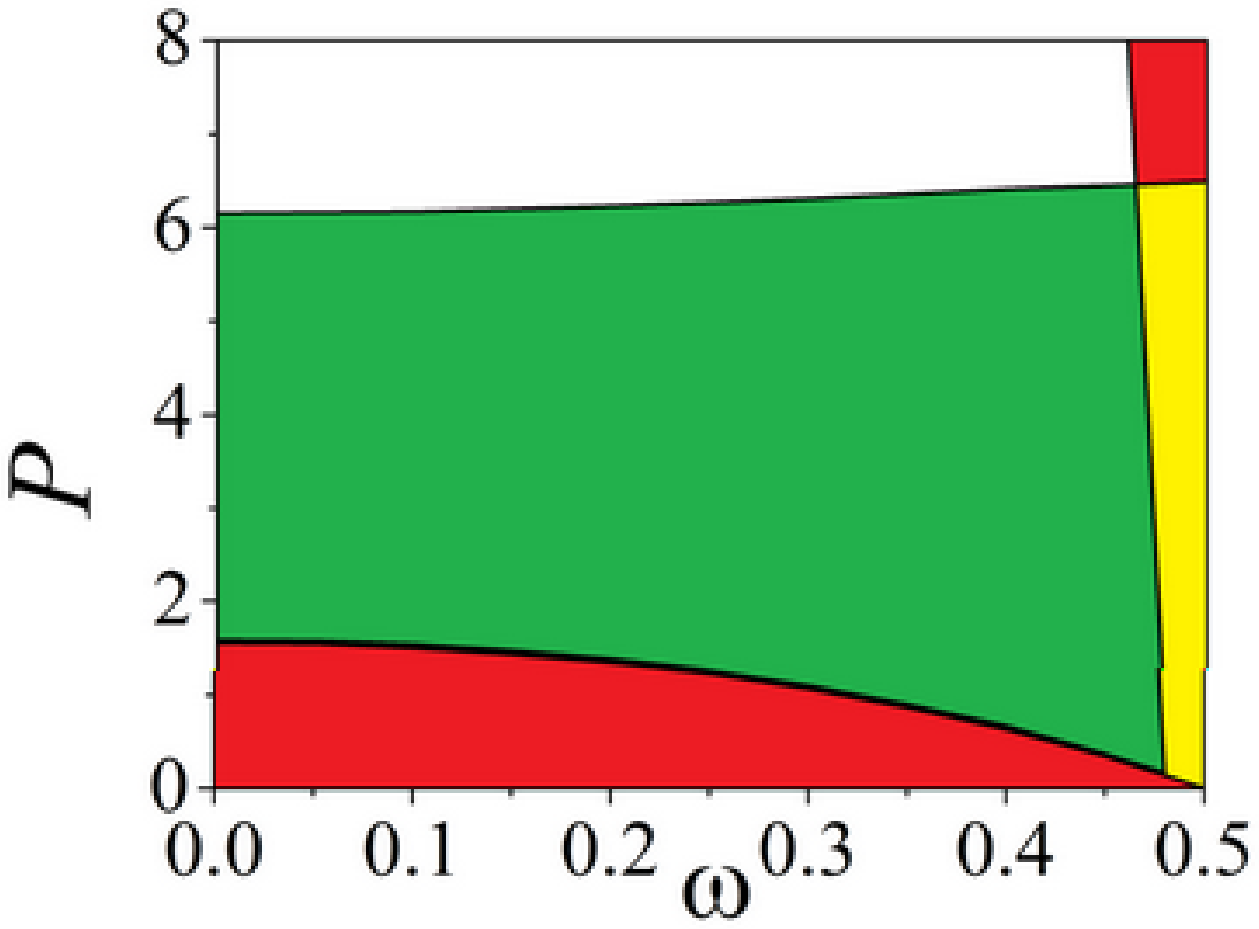}}
\caption{(Color online) (a) and (b): {Stability diagrams for the
even and odd (antisymmetric) modes, respectively, obtained in the
system with of the
self-focusing nonlinearity, in the plane of the rotation speed ($\protect%
\omega $) and total power ($P$). (c) The stability diagram for the
set of the second-harmonic (2H) and 2H-breaking modes. In panels
(a), (b), and (c), respectively, the even, odd, and 2H modes are
stable in the red areas, and unstable in the blank ones. In panel
(c), the 2H-breaking mode is stable in the green (middle) area, and
the bistability, i.e., coexistence of 2H and 2H-breaking stable
modes, occurs in the yellow (right edge) region. In the blank area
of panel (c), no 2H-breaking mode, stable or unstable one, is
found.}} \label{SFstab}
\end{figure}

\begin{figure}[tbp]
\centering%
\subfigure[] {\label{fig_55_a}
\includegraphics[scale=0.2]{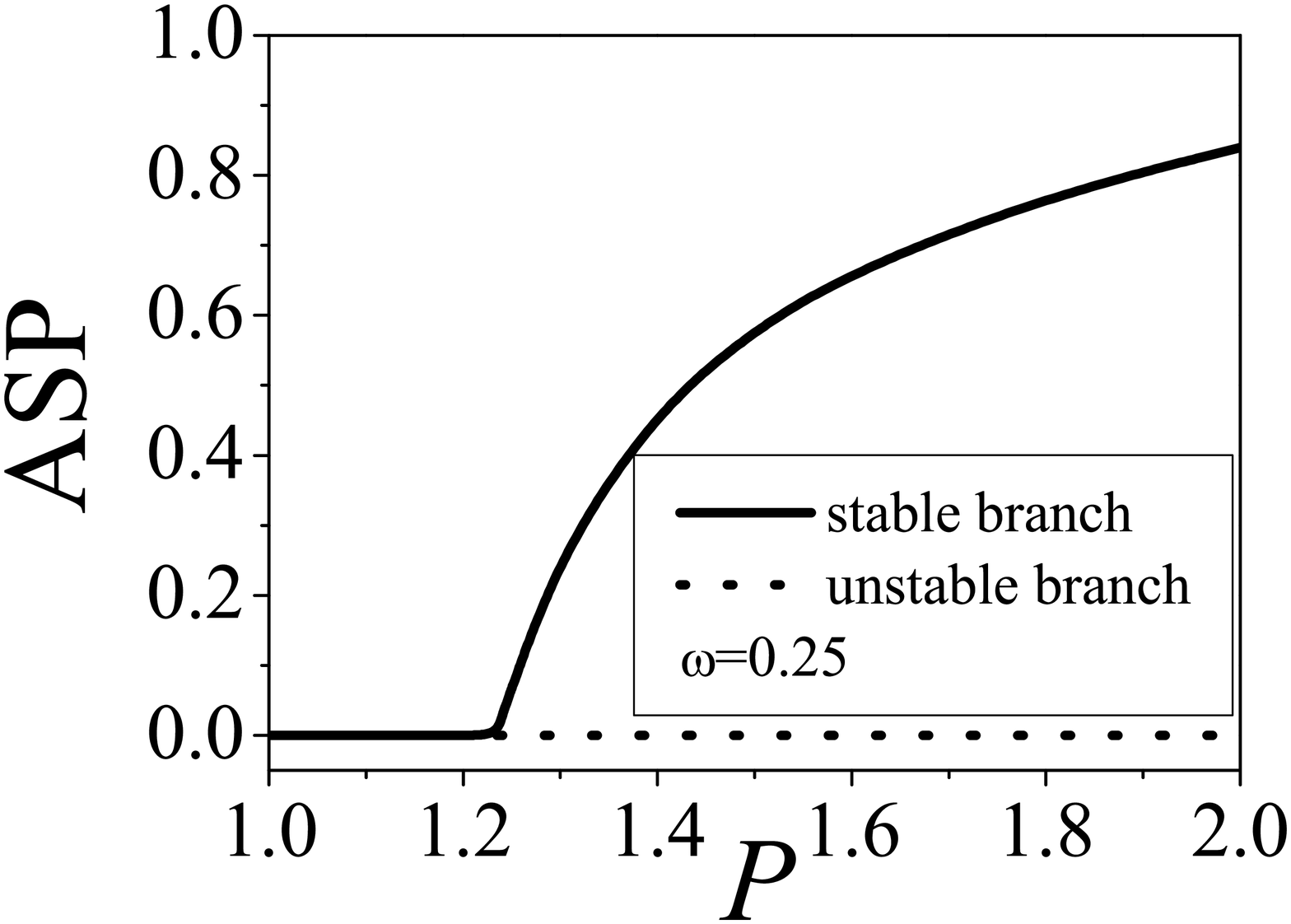}}%
\subfigure[] {\label{fig_55_b}
\includegraphics[scale=0.2]{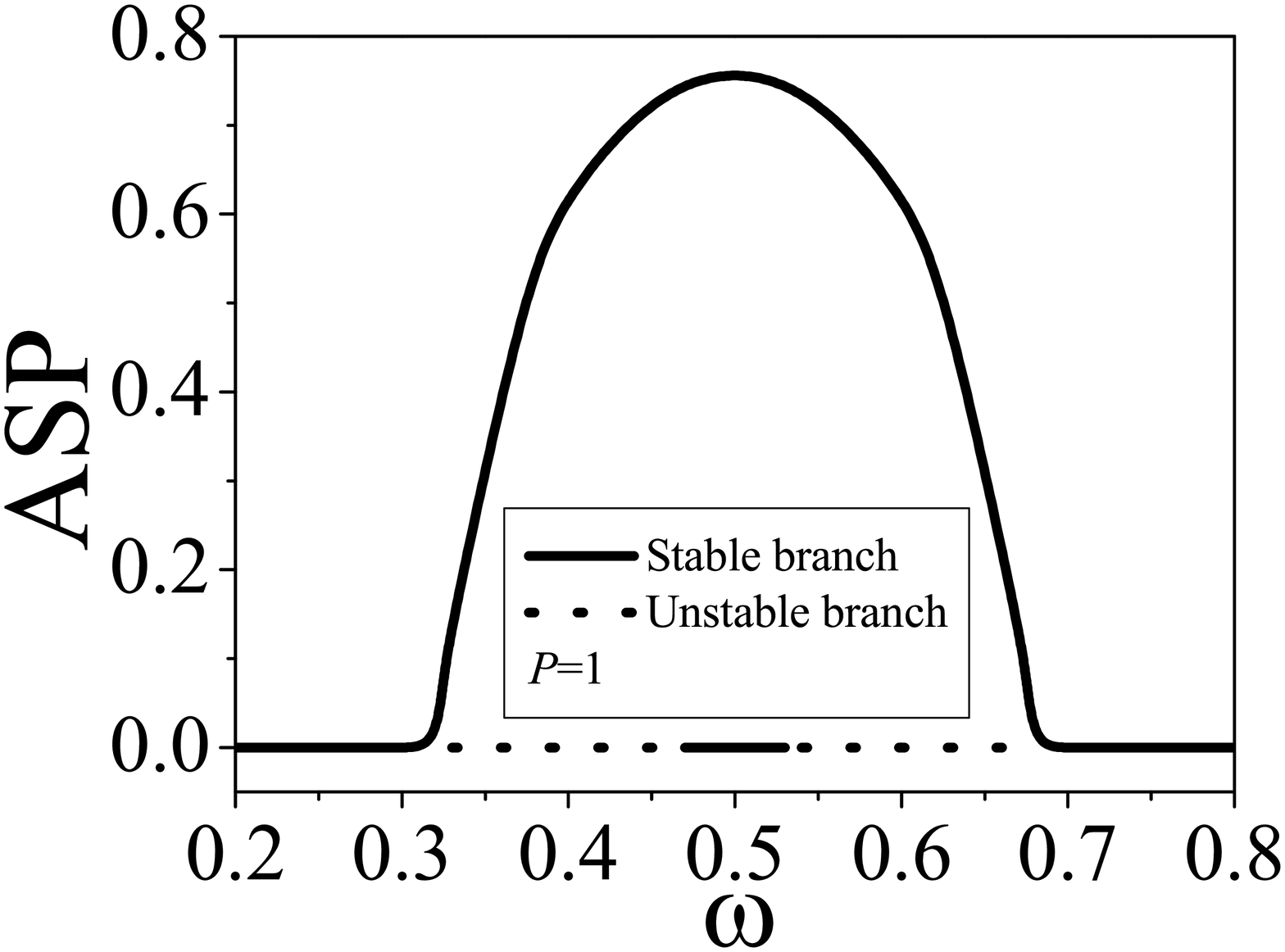}}
\subfigure[]{ \label{fig_55_c}
\includegraphics[scale=0.2]{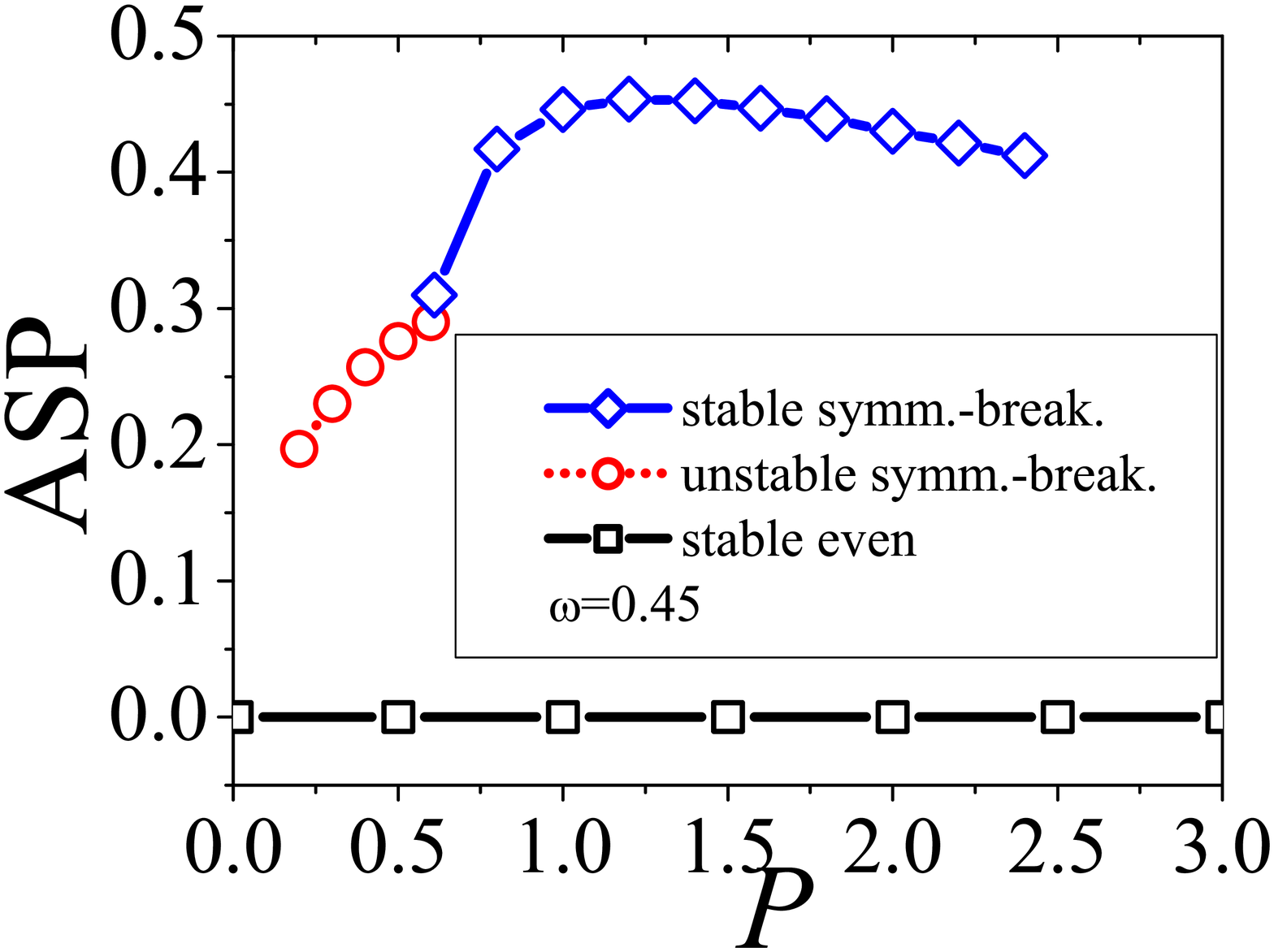}}
\caption{(Color online) (a)The ASP of the 2H and 2H-breaking modes, defined
as per Eq. (\protect\ref{measure}), as a function of the total power, $P$,
with the rotation speed fixed at $\protect\omega =0.25$, in the system with
the self-focusing nonlinearity (b). The same as a function of $\protect%
\omega $, at a fixed total power, $P=1$. The plot includes regions symmetric
with respect to $\protect\omega =0.5$, to stress the respective symmetry of
modes in the present system. (c) The ASP of even and symmetry-breaking mode
as a function of $P$ for $\protect\omega =0.45$, in the case of the
self-defocusing nonlinearity }
\label{ASP}
\end{figure}

Results of the numerical analysis for the stability of the modes in the
model with the SF nonlinear potential are summarized in Fig. \ref{SFstab},
in the form of diagrams drawn in the plane of ($P,\omega $). They show that
the stability regions of the even, odd, 2H and 2H-breaking modes
strongly overlap between themselves. In particular, the asymmetry parameter
of the 2H-breaking modes, which is defined as
\begin{equation}
\mathrm{ASP}\equiv \left\vert \int_{0}^{\pi }|\phi |^{2}d\theta ^{\prime
}-\int_{-\pi }^{0}|\phi |^{2}d\theta ^{\prime }\right\vert /P,  \label{measure}
\end{equation}
is displayed in Figs. \ref{fig_55_a} and \ref{fig_55_b}. The plots
demonstrate that the transition between 2H and 2H-breaking modes is of the
\textit{supercritical} type \cite{bif}.

\begin{figure}[tbp]
\centering\subfigure[] {\label{fig_5_a}
\includegraphics[scale=0.2]{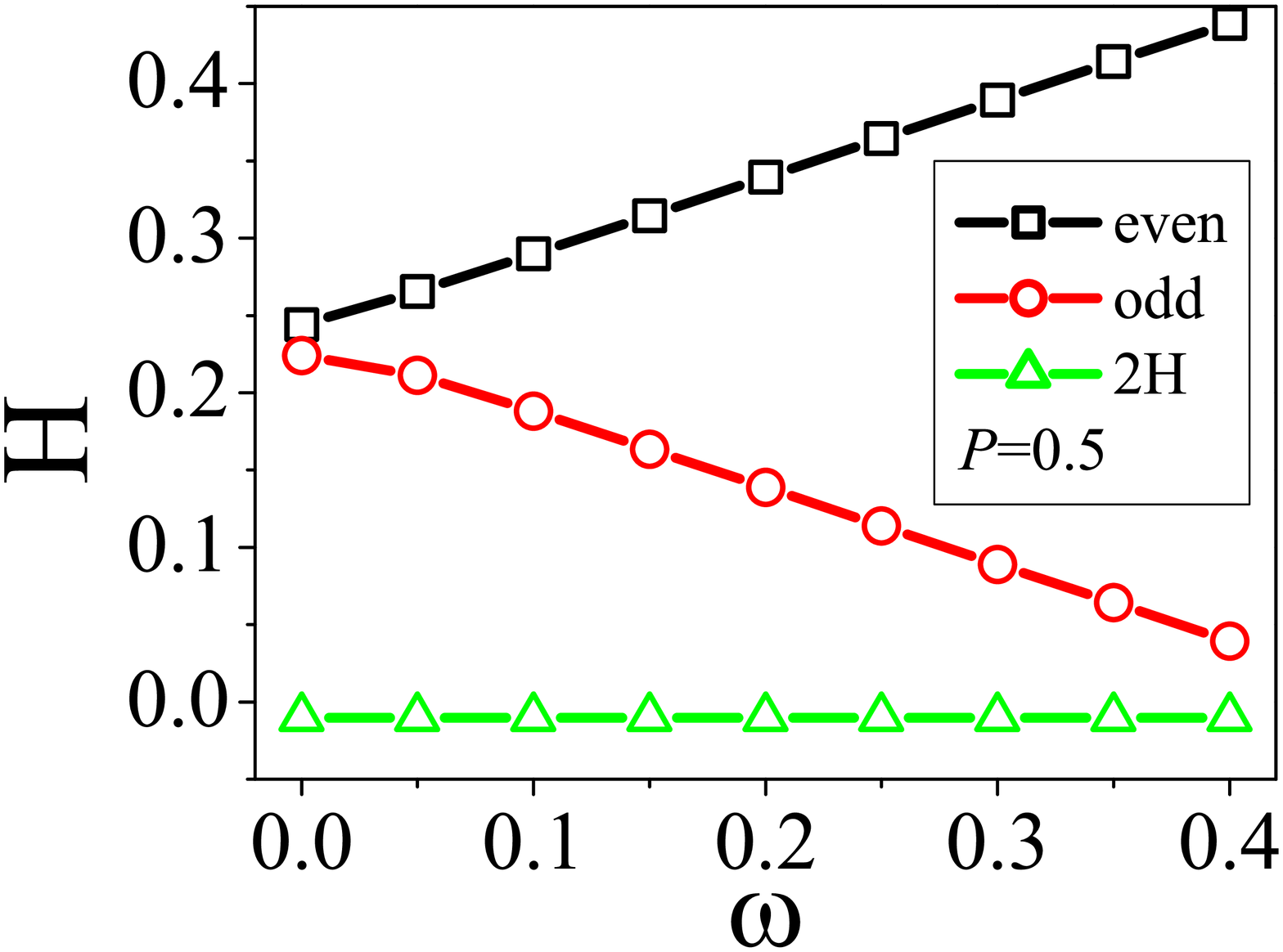}}%
\subfigure[] {\label{fig_5_b}
\includegraphics[scale=0.2]{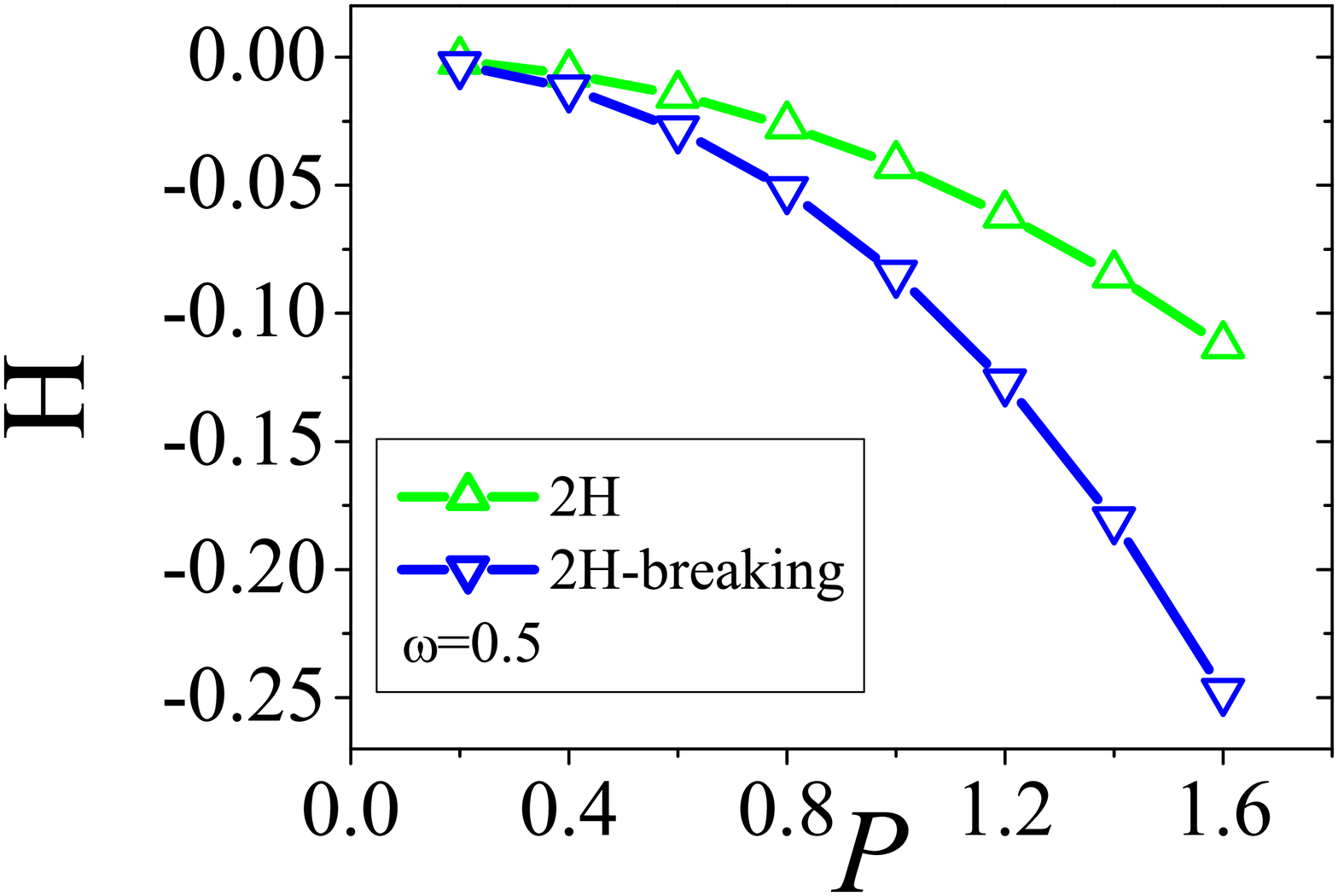}}
\subfigure[]{ \label{fig_5_c}
\includegraphics[scale=0.2]{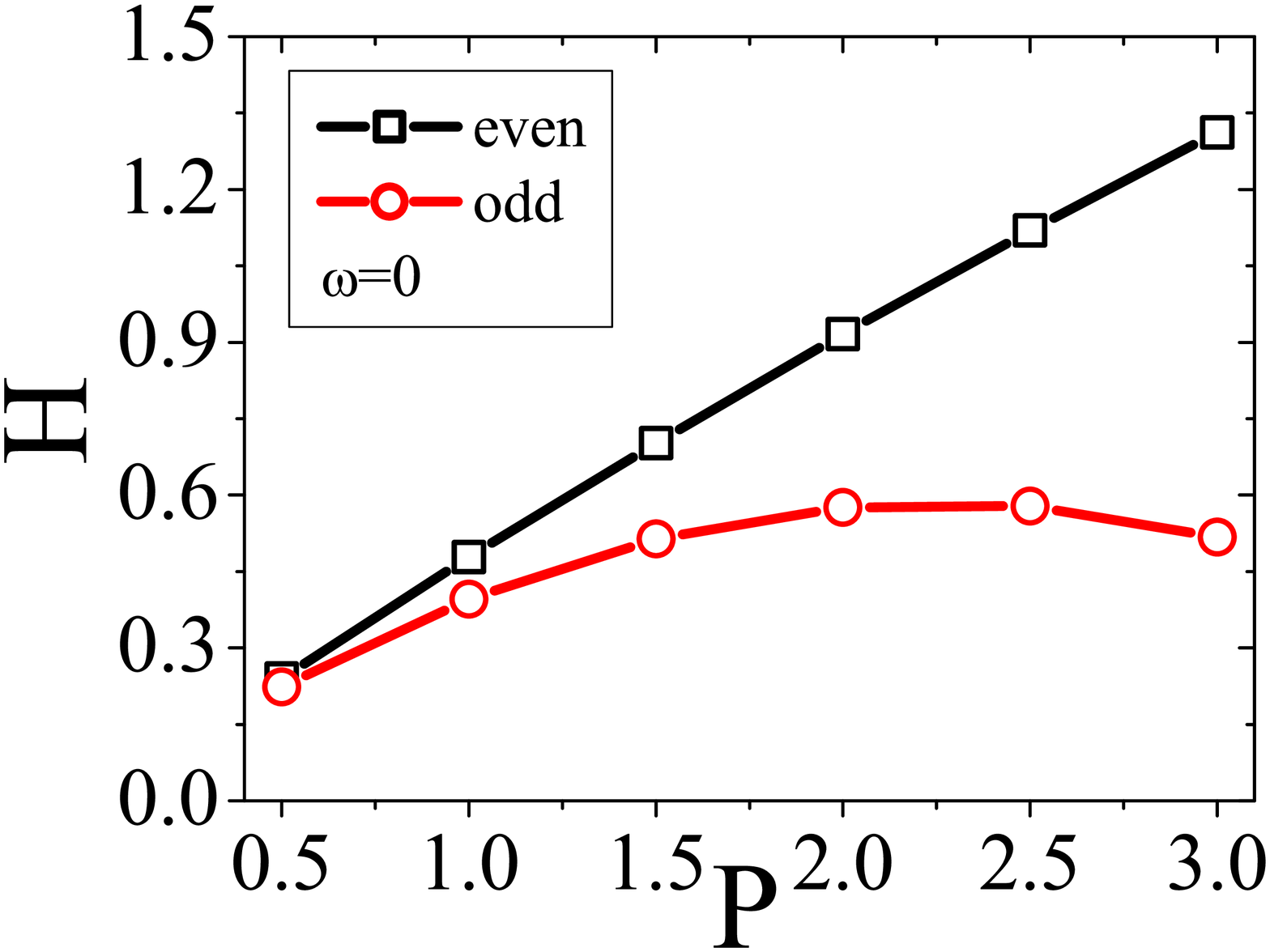}}
\caption{(Color online) (a) Energies of the even, odd, and second-harmonic
modes in the system with the self-focusing nonlinearity, computed, as per
Eq. (\protect\ref{Ham}), for $P=0.5$ and $0\leq \protect\omega \leq 0.4$.
(b) Energies of the 2H-breaking and second-harmonic modes, computed for $%
\protect\omega =0.5$ and $0.2\leq P\leq 1.6$. (c) Energies of the even and
odd modes along the vertical segment, with $\protect\omega =0$ and $0.5\leq
P\leq 3$. }
\label{SFHam}
\end{figure}

The multistability, which is obvious in Fig. \ref{SFstab}, makes it
necessary to compare energies of the coexisting dynamically stable modes,
defined by Eq. (\ref{Ham}), in order to identify the ground state that
realizes the energy minimum. First, in Fig. \ref{SFHam}(a) we show the
results along horizontal cuts of all the three panels of Fig. \ref{SFstab},
made at a constant value of the total power, $P=0.5$, with the rotation
speed varying in the interval of $0\leq \omega \leq 0.4$. The \textit{%
tristability} of the even, odd and 2H modes takes place along this segment.
Further, Fig. \ref{SFHam}(b) displays the results along vertical cuts of
panel (c) of Fig. \ref{SFstab} made at $\omega =0.5$, while the power is
varying as $0.2\leq P\leq 1.6$ [it is seen in panels (b) and (c) of Fig. \ref%
{SFstab} that the 2H-breaking mode also coexists with the stable odd one,
but the energy of the odd mode is definitely larger, therefore it is not
displayed in Fig. \ref{SFHam}(b)]. In panel (b) of Fig. \ref{SFHam}, the two
branches merge into one, with $H\rightarrow 0$, at $P\rightarrow 0$, as the
\emph{nonlinear} potential vanishes in this limit.

From Fig. \ref{fig_5_a}, we conclude that $H_{\mathrm{2H}}<H_{\mathrm{odd}%
}<H_{\mathrm{even}}$, while Fig. \ref{fig_5_b} shows that $H_{\mathrm{%
2H-break}}<H_{\mathrm{2H}}$. Calculations of the energy, performed along
other horizontal and vertical cuts, demonstrate that the following relation
between the energies of the different modes, suggested by these
inequalities, is always correct:
\begin{equation}
H_{\mathrm{2H-break}}<H_{\mathrm{2H}}<H_{\mathrm{odd}}<H_{\mathrm{even}}.
\label{H}
\end{equation}%
Thus, the 2H-breaking mode, if it exists [recall that, in the plane of $%
\left( \omega ,P\right) $ shown in Fig. \ref{SFstab}(b), it exists
above the curve separating the bottom (red) and middle (yellow)
areas], represents the ground state in the system with the
self-focusing nonlinearity. If the latter mode does not exist, then
the 2H state plays the same role.

In Fig. \ref{fig_5_a}, we can see that in the limit of $\omega =0$, the
energies of the even and odd modes are very close, which seems in
contradiction with the fact that these two modes are essentially different,
having the opposite parities. The reason is that the total power that we
chose here ($P=0.5$) is not large enough to show the energy difference
between the two modes. To clarify the point, in Fig. \ref{fig_5_c} we
display the energy curves for these two modes selected along the vertical
segment with $\omega =0$ and $0.5\leq P\leq 3$. It shows that the energy
difference indeed increases with the growth of $P$.

\subsection{The self-defocusing nonlinearity}

\begin{figure}[tbp]
\centering%
\subfigure[] {\label{fig_6_a}
\includegraphics[scale=0.21]{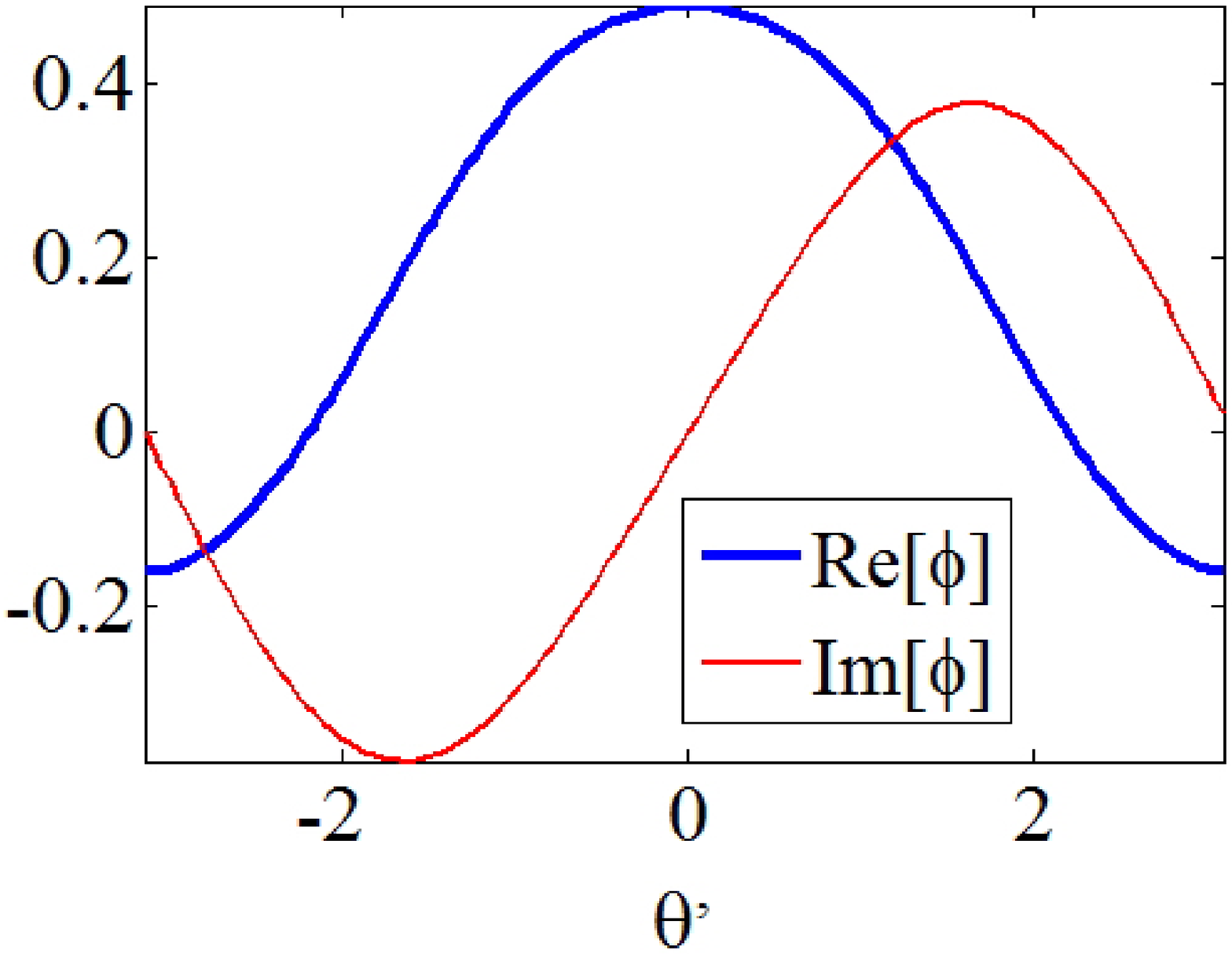}}%
\subfigure[] {\label{fig_6_b}
\includegraphics[scale=0.21]{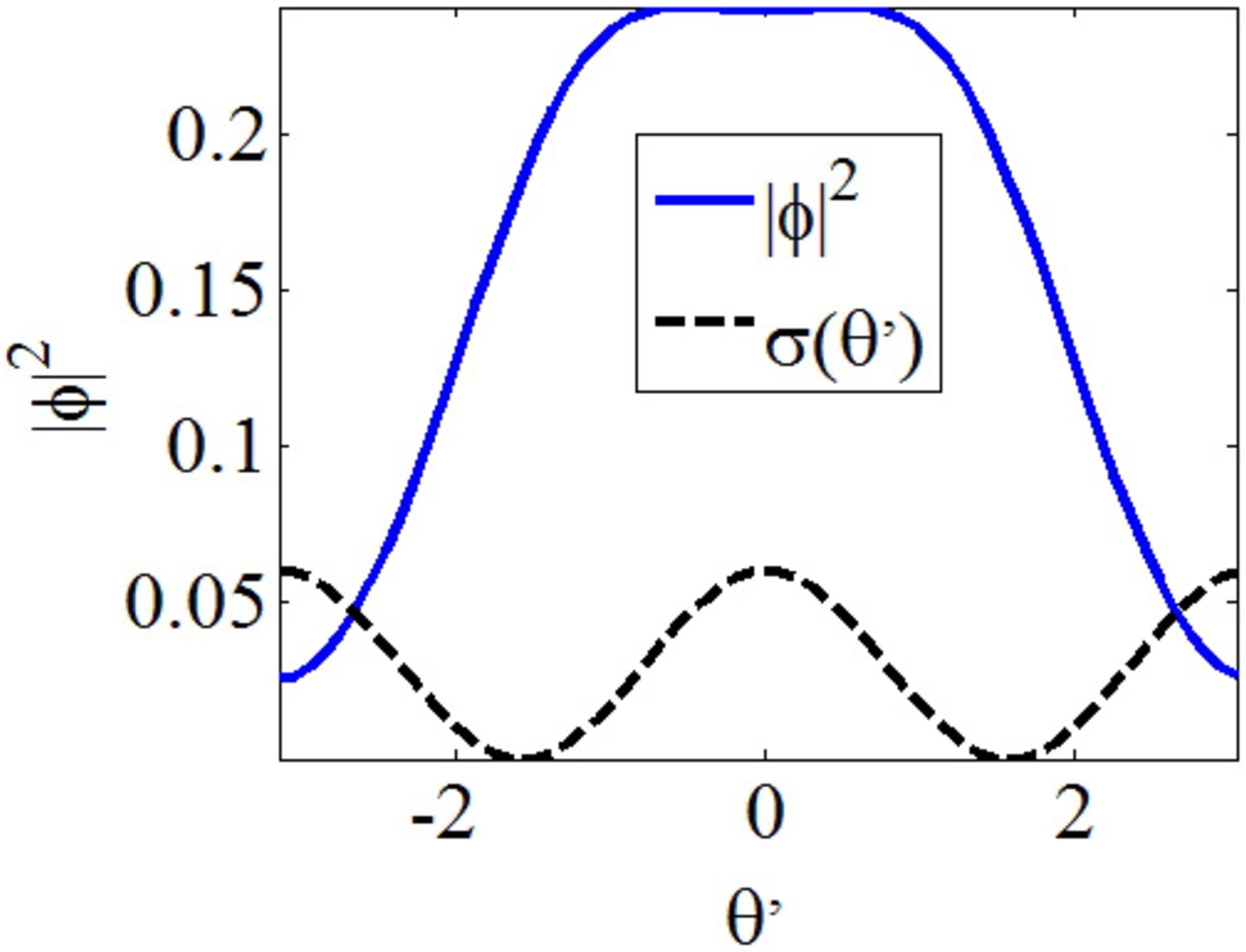}}
\caption{(Color online) A stable symmetry-breaking mode, found in the system
with the self-defocusing nonlinearity, at $(\protect\omega ,P)=(0.45,1)$:
(a) its real and imaginary parts; (b) the power profile.}
\label{SDFexp}
\end{figure}

In the case of the SDF nonlinear potential, represented by Eq. (\ref{SF-SDF}%
b), the numerical solution of Eq. (\ref{phi}) reveals stable modes of three
types, \textit{viz}., even, symmetry-breaking, and 2H (recall that
symmetry-breaking modes were not found in the system with the SF
nonlinearity), while odd and 2H-breaking states do not exist in this case.
Because the profiles of the even and 2H modes are quite similar to their
counterparts presented above in Fig. \ref{SFIP2H}, we here display, in Fig. %
\ref{SDFexp}, only a typical stable symmetry-breaking mode. The shape of
this mode seems symmetric, centered at $\theta ^{\prime }=0$; however, it is
classified as an asymmetric mode, as the power profile of a true symmetric
state would be double-humped, cf. Figs. \ref{SFIP2H}(b,d) and \ref{SFOP}(b),
while the present one has a single maximum, similar to the intensity
distribution in the 2H-breaking state in Fig. \ref{SFOP}(d). Accordingly,
the ASP (effective asymmetry measure) for the symmetry-breaking mode is
introduced as follows, instead of the above definition (\ref{measure}):%
\begin{equation}
\left( \mathrm{ASP}\right) _{\mathrm{symm-break}}\equiv \left\vert \left[
\int_{-\pi /2}^{+\pi /2}-\left( \int_{-\pi }^{-\pi /2}+\int_{+\pi /2}^{+\pi
}\right) \right] |\phi |^{2}d\theta ^{\prime }\right\vert /P,
\label{ASP-symmbr}
\end{equation}%
to stress the lack of the asymmetry between the central and peripheral parts
parts of the mode.

\begin{figure}[tbp]
\centering%
\subfigure[] {\label{fig_7_a}
\includegraphics[scale=0.42]{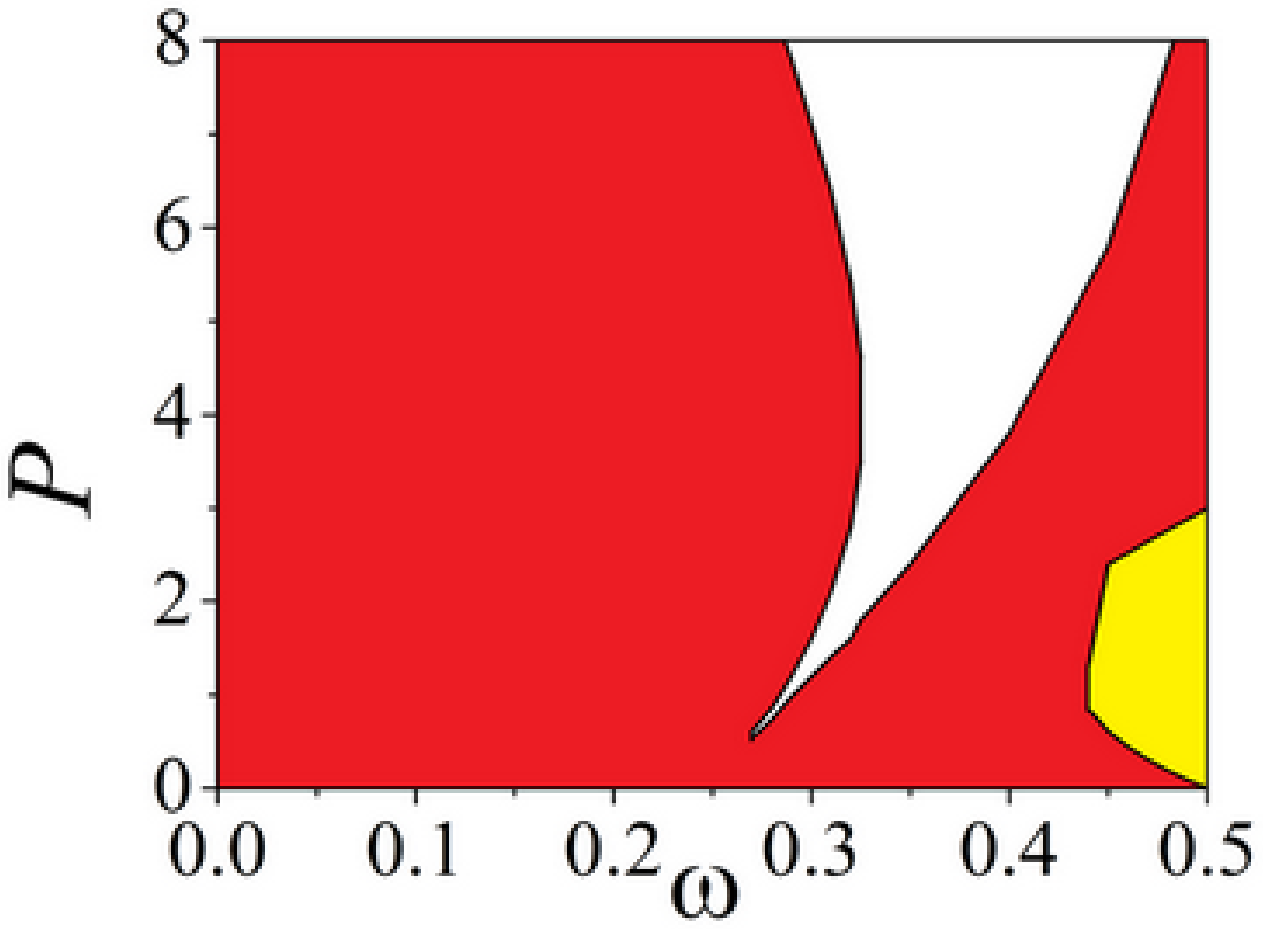}}%
\subfigure[] {\label{fig_7_b}
\includegraphics[scale=0.2]{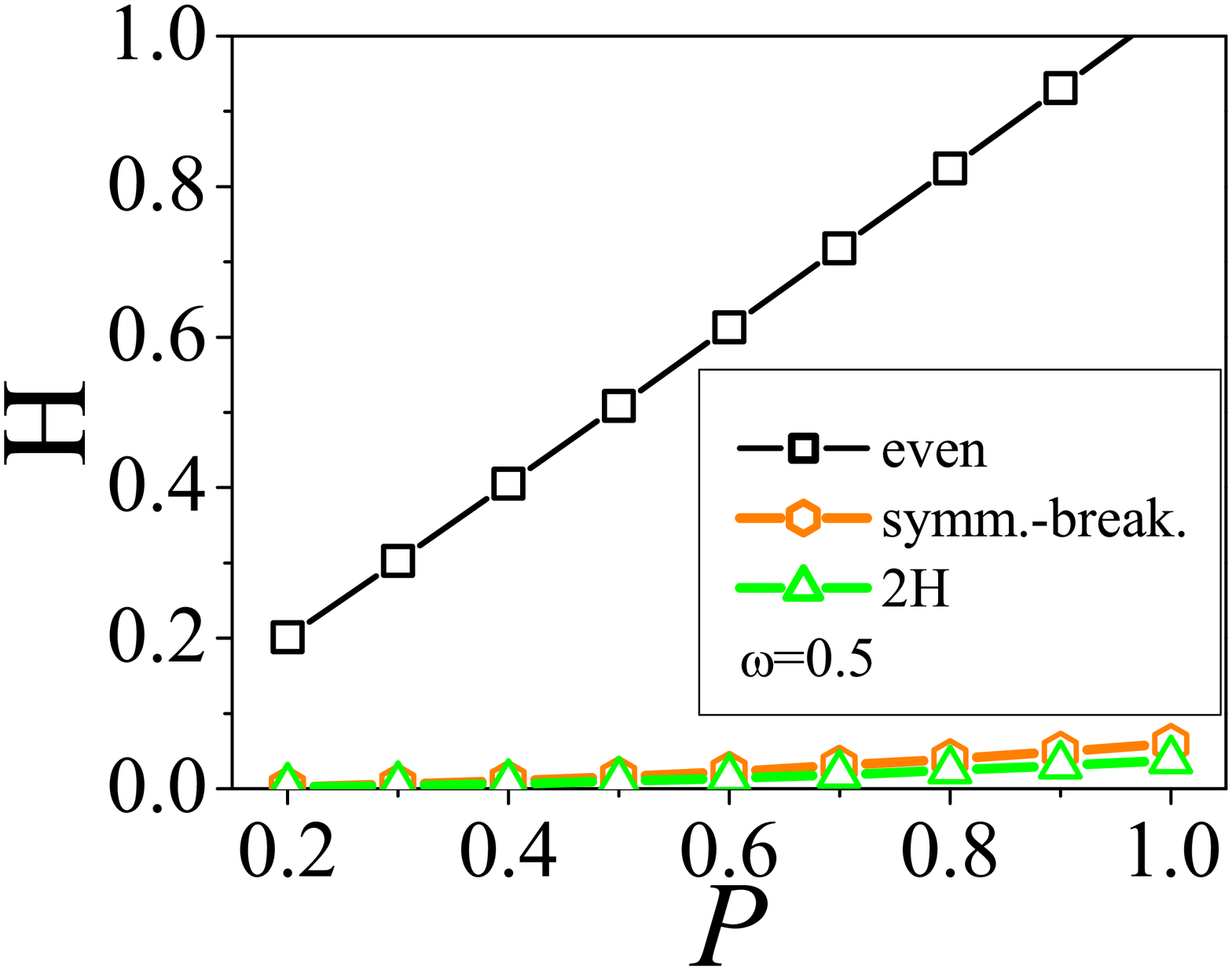}}
\caption{(Color online) (a) The stability diagram for the even and
symmetry-breaking modes in the system with the self-defocusing nonlinearity.
The red (largest) and yellow (smallest, near the right edge) areas
designate, severally, the stability region of the even mode, and the region
of the coexistence (bistability) of the even and symmetry-breaking modes. In
the blank area, no stable modes of these types are found (in fact, an
unstable even mode exists in that area). The second-harmonic mode exists and
is stable in the entire plane. (b) Energies of the even, symmetry-breaking,
and second-harmonic modes along the vertical cut of panel (a) at $\protect%
\omega =0.5$ and $0.2\leq P\leq 1$.}
\label{SDFIPA}
\end{figure}

The stability and energy diagrams for the even, symmetry-breaking, and 2H
modes are displayed in Fig. \ref{SDFIPA} [the 2H mode exists and is stable
in the entire plane of $(P,\omega )$, therefore it is not specially marked
in panel \ref{SDFIPA}(a)]. In particular, it is observed that the
symmetry-breaking mode exists near the right edge of the rotational
Brillouin zone, i.e., it does not exist in the stationary system (with $%
\omega =0$). The ASP of the even and symmetry-breaking modes, defined per
Eq. (\ref{ASP-symmbr}), is displayed in Fig. \ref{fig_55_c} as a function of
of the total power. The absence of the linkage between the branches
representing these two modes implies that they are not related by any
bifurcation.

Figure \ref{fig_7_b} displays the comparison of energies of these three
kinds of the modes (even, symmetry-breaking, and 2H ones) along the vertical
cut made at $\omega =0.5$, with the power varying in interval $0.2\leq P\leq
1$. It demonstrates that the curve for the 2H mode goes close to but
slightly lower than its counterpart for the symmetry-breaking mode. The
analysis of more general data, produced by the numerical calculations for
the system with the SDF nonlinear potential, demonstrates that energies of
all the three dynamically stable modes existing in this case are ordered as
follows:
\begin{equation}
H_{\mathrm{2H}}<H_{\mathrm{symm-break}}<H_{\mathrm{even}},  \label{H2}
\end{equation}%
cf. Eq. (\ref{H}). Thus, the 2H mode plays the role of the ground state in
the case of the SDF nonlinearity (recall this mode exists at all values of $%
\omega $ and $P$).

\subsection{The alternating self-focusing -- self-defocusing nonlinear
potential}

\begin{figure}[tbp]
\centering%
\subfigure[] {\label{fig_8_a}
\includegraphics[scale=0.18]{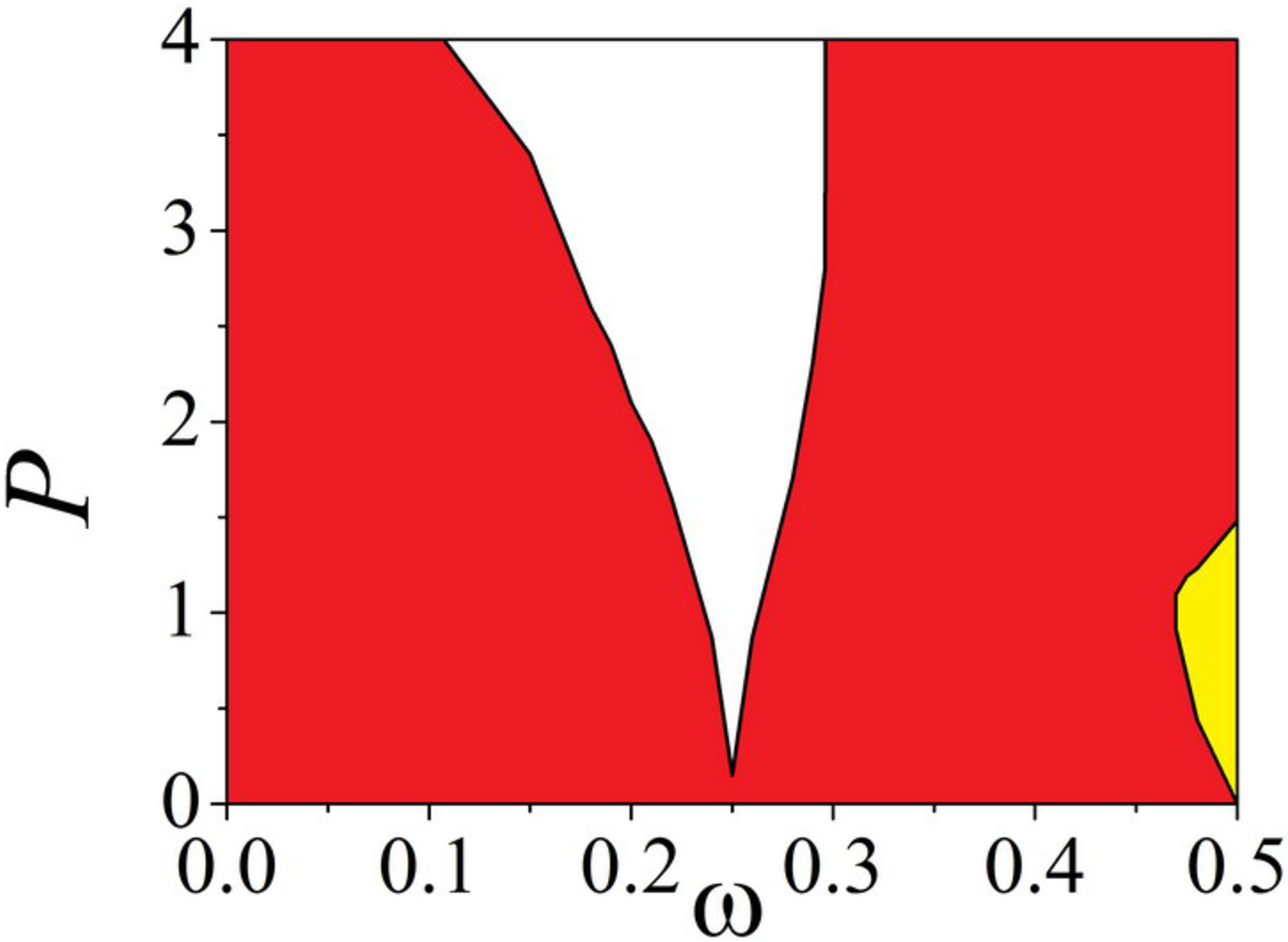}}%
\subfigure[] {\label{fig_8_b}
\includegraphics[scale=0.35]{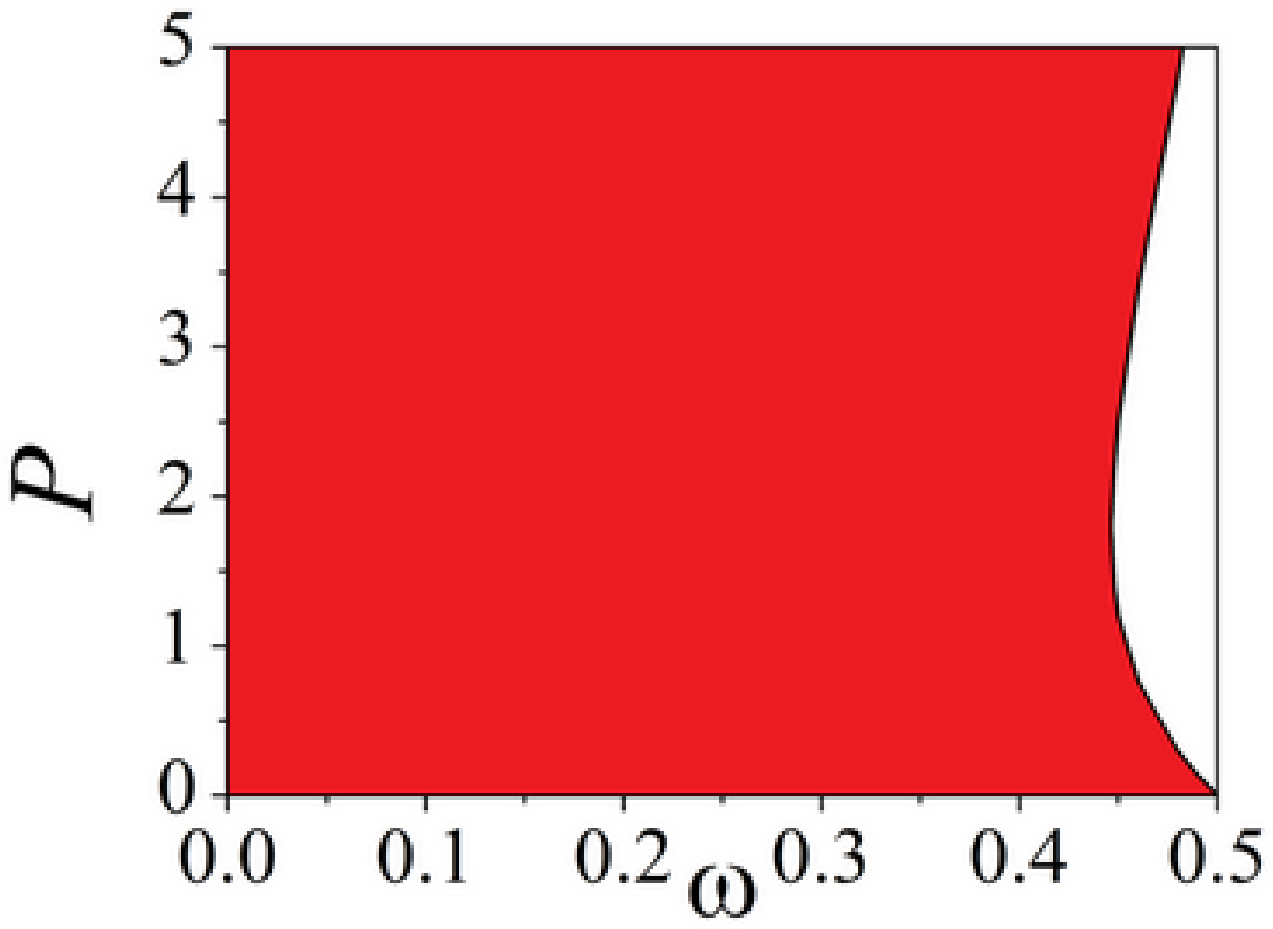}}
\subfigure[]{ \label{fig_8_c}
\includegraphics[scale=0.35]{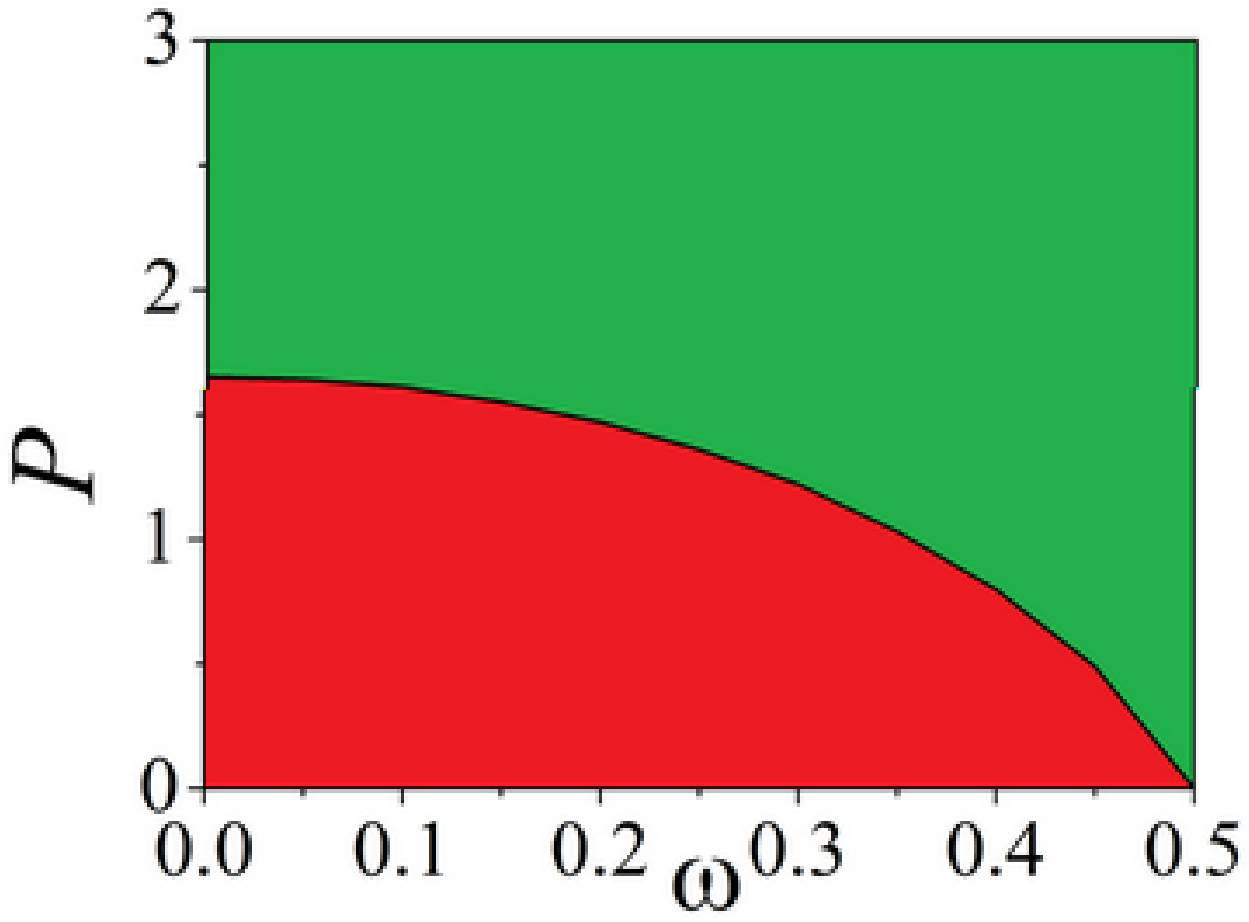}}
\caption{(Color online) Stability diagrams for the system with the
alternating self-focusing -- self-defocusing nonlinearity. (a) Even and
symmetry-breaking modes; (b) odd and 2H-breaking modes; (c) the 2H
modes. In (a), the even mode is stable in the red (bottom) area; in the
small yellow region it coexists with the symmetry-breaking one, and no
stable modes of these types are found in the blank area. In (b), the odd
mode is stable in the red (bottom) area, the 2H-breaking mode is
stable in the green region (adjacent to the right edge of the panel), and
these two types coexist in the yellow area. In (c), the 2H mode is stable in
the red area.}
\label{SFSDF}
\end{figure}

In the case of the alternating SF-SDF nonlinearity, defined as per Eq. (\ref%
{SF-SDF}), the numerical solutions reveal the existence of all the five
types of stable trapped modes indicated in Table \ref{tab:guess}. Profiles
of these modes are quite similar to those of their counterparts displayed
above in Figs. \ref{SFIP2H}, \ref{SFOP}, and \ref{SDFexp}, therefore we do
not show them again here. The respective stability diagrams in the $%
(P,\omega )$ plane are presented in Fig. \ref{SFSDF}. In particular, the
absence of a bistability area in panel (c) of Fig. \ref{SFSDF} suggests that
the transition between the 2H and 2H-breaking modes is supercritical, like
in the system with the SF nonlinearity, cf. Fig. \ref{fig_4_c}.

The comparison of the energies of the five types of the modes which may be
stable in the case of the sign-alternating nonlinearity demonstrates the
following ordering, cf. Eqs. (\ref{H}) and (\ref{H2}):
\begin{equation}
H_{\mathrm{2H-break}}<H_{\mathrm{2H}}<H_{\mathrm{symm.-break}},~H_{\mathrm{%
odd}}<H_{\mathrm{even}}.  \label{H3}
\end{equation}%
The energies of the symmetry-breaking and odd modes are not compared in Eq. (%
\ref{H3}), as their stability regions do not overlap, see Fig. \ref{SFSDF}%
(a,b). Thus, the 2H-breaking mode, when it exists, plays the role of the
ground state in the present case; otherwise, the ground state is represented
by the 2H mode, see panels (b) and (c) in Fig. \ref{SFSDF}.

\section{The analytical approach}

\subsection{The two-mode approximation}

The present setting may be naturally approximated by a finite-mode
truncation of the expansion of stationary field $\phi (x)$ over the set of
spatial harmonics. The simplest approximation reduces to the substitution of
truncation%
\begin{equation}
\phi \left( \theta ^{\prime }\right) =a_{0}+a_{1}\exp \left( i\theta
^{\prime }\right)  \label{ansatz}
\end{equation}%
into Eq. (\ref{phi}). In this expression, $a_{0}$ may be fixed to be real,
while amplitude $a_{1}$ is allowed to be complex. This approach is
consistent if, in the linear approximation, each term in combination (\ref%
{ansatz}) is an exact solution of Eq. (\ref{phi}) for a common value of $\mu
$, hence the zeroth-order approximation exists, and the analysis of weakly
nonlinear states can be developed around it. It is easy to see that such a
case corresponds, in the zeroth approximation, to $\omega =1/2$ [which is
exactly the right edge of zone (\ref{zone})] and $\mu =0$ \cite{we}. Then,
weakly nonlinear modes can be constructed in an approximate analytical form,
assuming that $\mu $, $a_{0}^{2},$ $\left\vert a_{1}\right\vert ^{2}$, and
\begin{equation}
\delta \equiv 1/2-\omega  \label{delta}
\end{equation}%
are all small quantities. To this end, ansatz (\ref{ansatz}) and a
particular expression (\ref{SF-SDF}) for $\sigma \left( \theta ^{\prime
}\right) $ are substituted into Eq. (\ref{phi}), and equations for
amplitudes $a_{0}$ and $a_{1}$ are derived as balance conditions for the
zeroth and first harmonics.$\allowbreak $

Ansatz (\ref{ansatz}) corresponds to the following approximation for the
total power (\ref{Power}),
\begin{equation}
P=\int_{-\pi }^{+\pi }|\phi |^{2}d\theta ^{\prime }=2\pi
(a_{0}^{2}+\left\vert a_{1}\right\vert ^{2}),  \label{P}
\end{equation}%
which will be used below too.

\subsection{The self-focusing nonlinearity}

In the case of the SF nonlinear potential (\ref{SF-SDF}a), the two-mode
approximation (\ref{ansatz}) leads to the following equations:%
\begin{eqnarray}
&&\mu =-{\frac{1}{2}}a_{0}^{2}-|a_{1}|^{2}+\frac{1}{4}a_{1}^{2},  \notag \\
&&\mu -\delta =-a_{0}^{2}-{\frac{1}{2}}|a_{1}|^{2}+{\frac{1}{4}}a_{0}^{2}{%
\frac{a_{1}^{\ast }}{a_{1}}}.  \label{SFa0a1}
\end{eqnarray}%
As shown above by the numerical analysis, the SF nonlinearity gives rise,
\textit{inter alia}, to the 2H-breaking mode, which is generated from input $%
b+\sin \theta ^{\prime }$ in Table \ref{tab:guess}. To capture the part of
the solution corresponding to $\sin \theta ^{\prime }$ in the input, we set
\begin{equation}
a_{1}=ic,  \label{i}
\end{equation}%
Then, Eq. (\ref{SFa0a1}) yields
\begin{eqnarray}
&&a_{0}^{2}=\left( {4/21}\right) (5\delta -3\mu ),  \notag \\
&&c^{2}=-\left( {4/21}\right) (2\delta +3\mu ).  \label{SFac}
\end{eqnarray}%
Substituting solution (\ref{SFac}) into expression (\ref{P}), we obtain a
relation between the propagation constant and total power for the
2H-breaking mode,
\begin{equation}
\mu ={\frac{\delta }{2}}-{\frac{7}{16\pi }}P.  \label{SF_EP}
\end{equation}%
It demonstrates that this mode, as predicted by the analytical
approximation, satisfied the Vakhitov-Kolokolov (VK) criterion, $d\mu /dP<0$%
, which is a necessary stability condition for patterns supported by the SF
nonlinearity \cite{VK}.

Furthermore, Eq. (\ref{SFac}) predicts the emergence of the 2H-breaking
modes at $a_{0}^{2}=0$, i.e.,
\begin{equation}
\mu =5\delta /3.  \label{5/3}
\end{equation}%
Since the modes with rotation speeds related by Eq. (\ref{+-}) are mutually
tantamount, Eqs. (\ref{5/3}) and (\ref{SF_EP}) predict the coexistence of
the odd and 2H-breaking modes at
\begin{equation}
P\geq \left( P_{\mathrm{\min }}\right) =\left( 8\pi /3\right) |\delta |.
\label{SFborder}
\end{equation}%
This analytical result is compared with the corresponding numerical findings
in Fig. \ref{Comparison}(a), which shows a reasonably good agreement.

\subsection{The self-defocusing nonlinearity}

The substitution of the same ansatz (\ref{ansatz}) into Eq. (\ref{phi}) in
the case of the SDF nonlinear potential, taken as per Eq. (\ref{SF-SDF}b),
yields the following algebraic equations, instead of Eq. (\ref{SFa0a1})
derived above for the SF nonlinearity:
\begin{eqnarray}
&&\mu ={\frac{1}{2}}a_{0}^{2}+|a_{1}|^{2}+{\frac{1}{4}}a_{1}^{2},  \notag \\
&&\mu -\delta =a_{0}^{2}+{\frac{1}{2}}|a_{1}|^{2}+{\frac{1}{4}}a_{0}^{2}{%
\frac{a_{1}^{\ast }}{a_{1}}}.  \label{SFa0a2}
\end{eqnarray}%
The above numerical results for the SDF nonlinearity demonstrate the
existence of the symmetry-breaking mode in this case, which is generated by
input $b+\cos \theta ^{\prime }$ in Table \ref{tab:guess}. To capture this
mode by means of ansatz (\ref{ansatz}), it is natural to set
\begin{equation}
a_{1}=c\equiv \mathrm{\ real},  \label{r}
\end{equation}%
on the contrary to Eq. (\ref{i}), where $a_{1}$ was imaginary. In this case,
the algebraic system (\ref{SFa0a2}) yields%
\begin{eqnarray}
&&a_{0}^{2}=\left( {4/21}\right) (3\mu -5\delta ),  \notag \\
&&c^{2}=\left( {4/21}\right) (3\mu +2\delta ),  \label{SDFac}
\end{eqnarray}%
cf. Eq. (\ref{SFac}). According to Eq. (\ref{P}), the relations between the
propagation constant and total power takes the following form for solution (%
\ref{SDFac}):
\begin{equation}
\mu ={\frac{\delta }{2}}+{\frac{7}{16\pi }}P,  \label{anti}
\end{equation}%
cf. Eq. (\ref{SF_EP}). This equation demonstrates that the asymmetric mode
satisfies the \textit{anti-VK} criterion, $d\mu /dP>0$, which, as argued in
Ref. \cite{anti}, may play the role of a necessary stability condition for
modes supported by the SDF nonlinearity.

The analytical approximation predicts the existence boundary for the
symmetry-breaking states, $a_{0}^{2}=0$, in the same form (\ref{5/3}) as it
was obtained above for the SF nonlinearity. Consequently, the existence
region for these states is predicted in the form coinciding with that given
by Eq. (\ref{SFborder}). This result is compared with its numerical
counterpart in Fig. \ref{Comparison}(b).

\subsection{The alternating self-focusing -- self-defocusing nonlinearity}

In the case of the alternating SF-SDF nonlinear potential (\ref{SF-SDF}c),
the substitution of ansatz (\ref{ansatz}) into Eq. (\ref{phi}) yields the
algebraic equations in the form which is somewhat simpler than Eqs. (\ref%
{SFa0a1}) and (\ref{SFa0a2}) derived above for the ``pure" SF and SDF
nonlinearities:%
\begin{eqnarray}
&&\mu ={\frac{1}{2}}a_{1}^{2},  \notag \\
&&\mu -\delta ={\frac{1}{2}}a_{0}^{2}{\frac{a_{1}^{\ast }}{a_{1}}}.
\label{SFSDFa0a1}
\end{eqnarray}%
First, if we choose real $a_{1}$, as in Eq. (\ref{r}), which refers to the
symmetry-breaking mode, Eq. (\ref{SFSDFa0a1}) yields
\begin{eqnarray}
&&a_{0}^{2}=2(\mu -\delta ),  \notag \\
&&c^{2}=2\mu .  \label{SFSDFac}
\end{eqnarray}

\begin{figure}[tbp]
\centering%
\subfigure[] {\label{fig_9_a}
\includegraphics[scale=0.42]{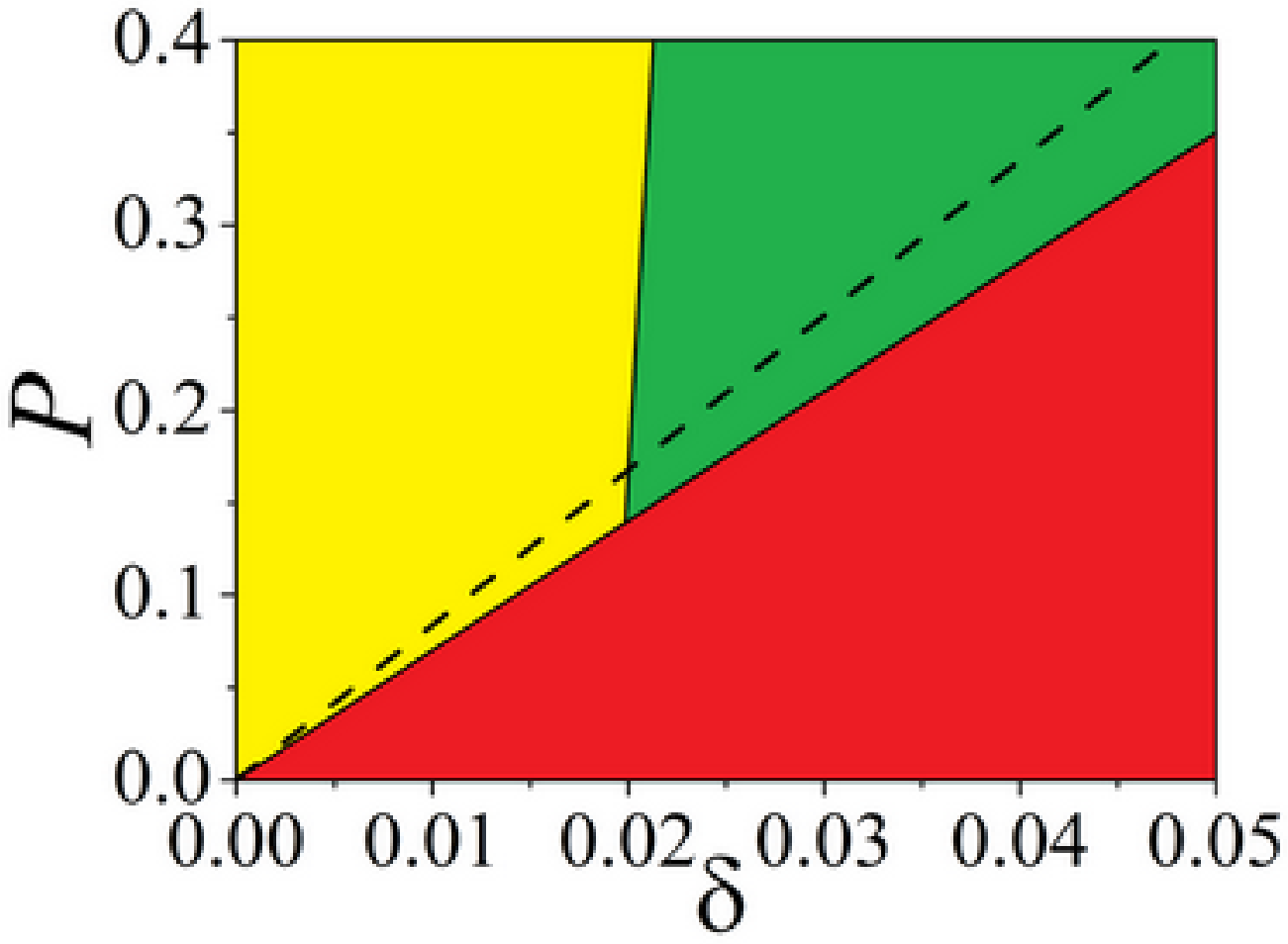}}%
\subfigure[] {\label{fig_9_b}
\includegraphics[scale=0.42]{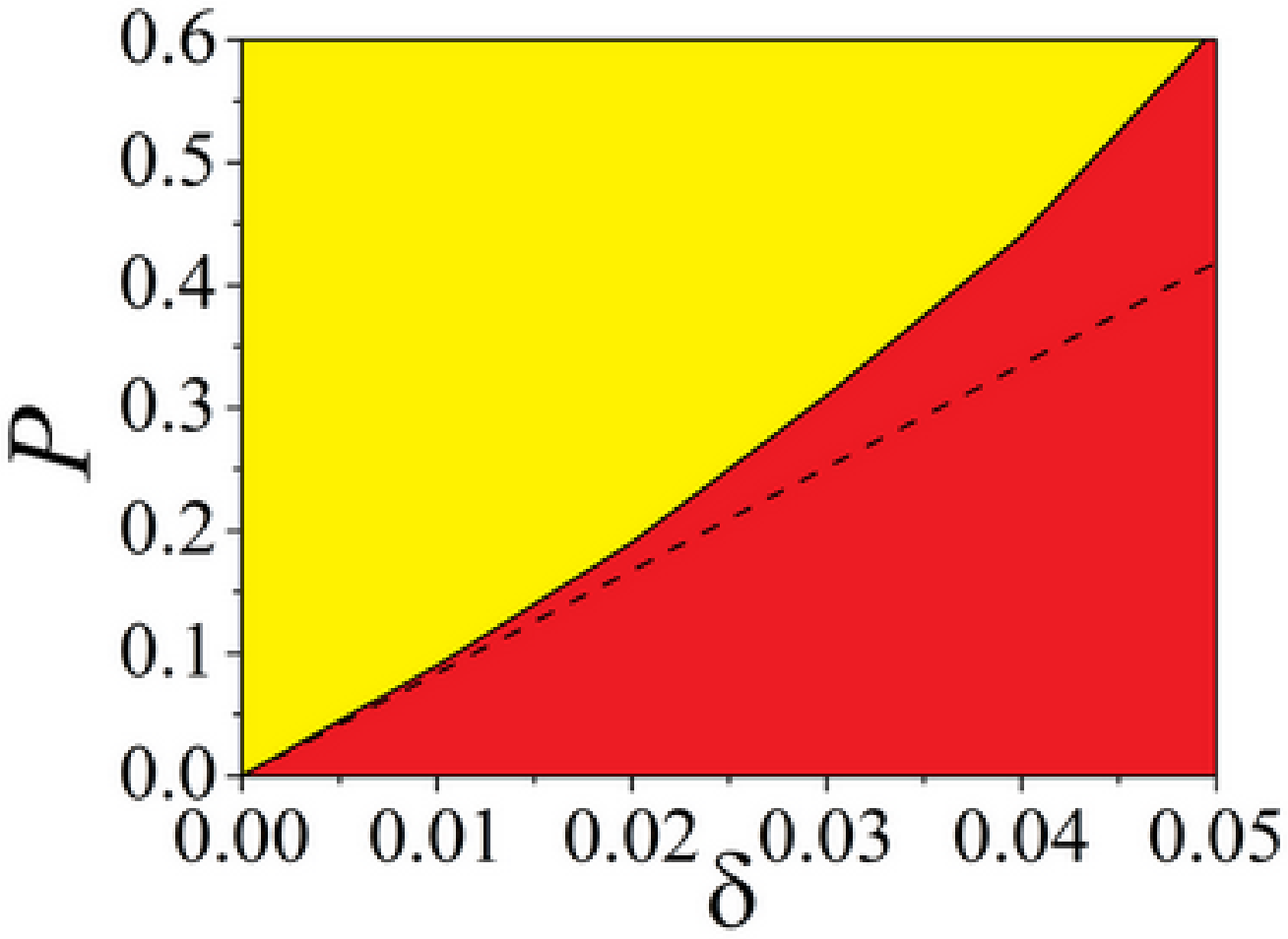}}
\subfigure[]{ \label{fig_9_c}
\includegraphics[scale=0.42]{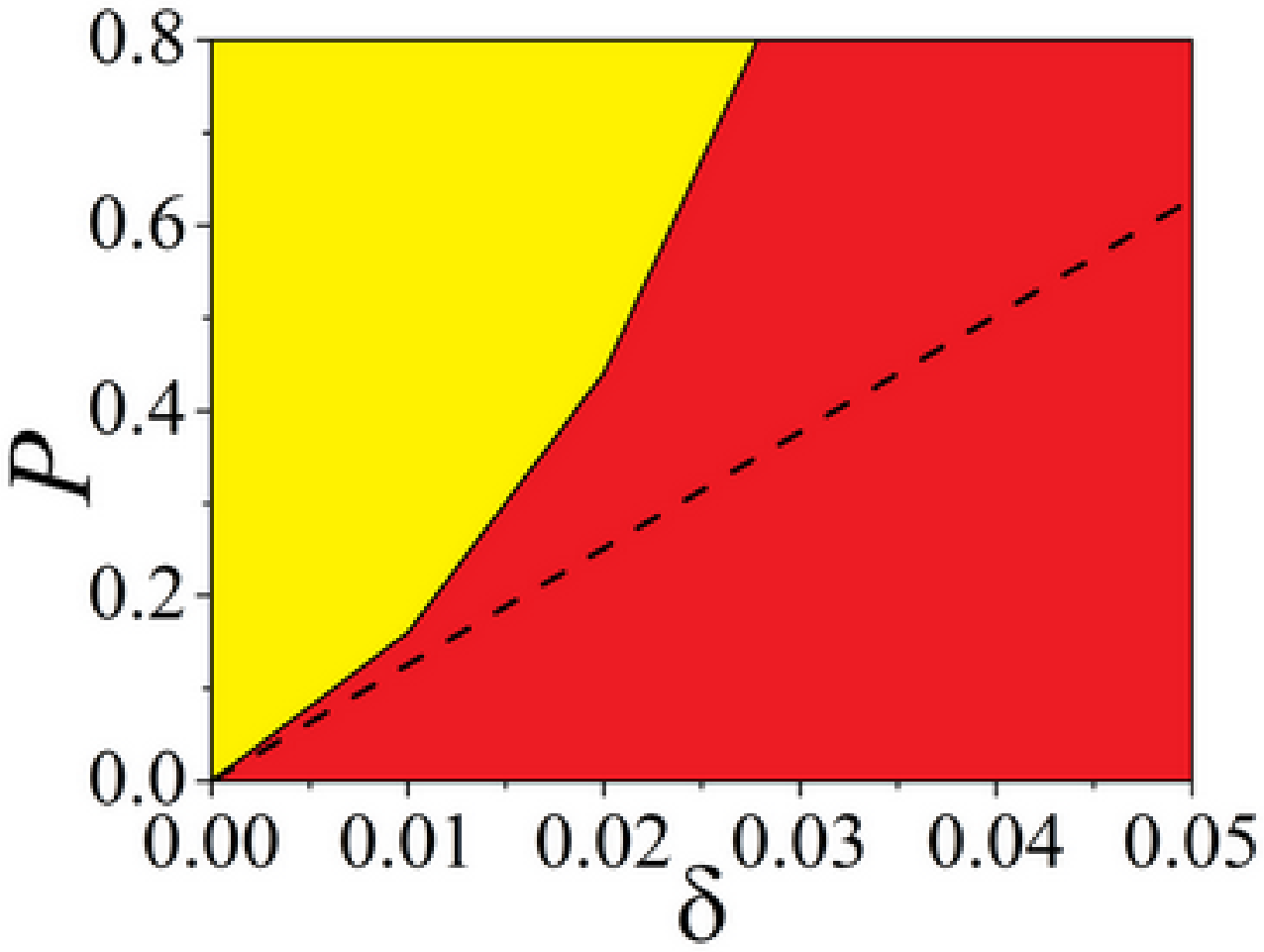}}
\subfigure[]{ \label{fig_9_d}
\includegraphics[scale=0.42]{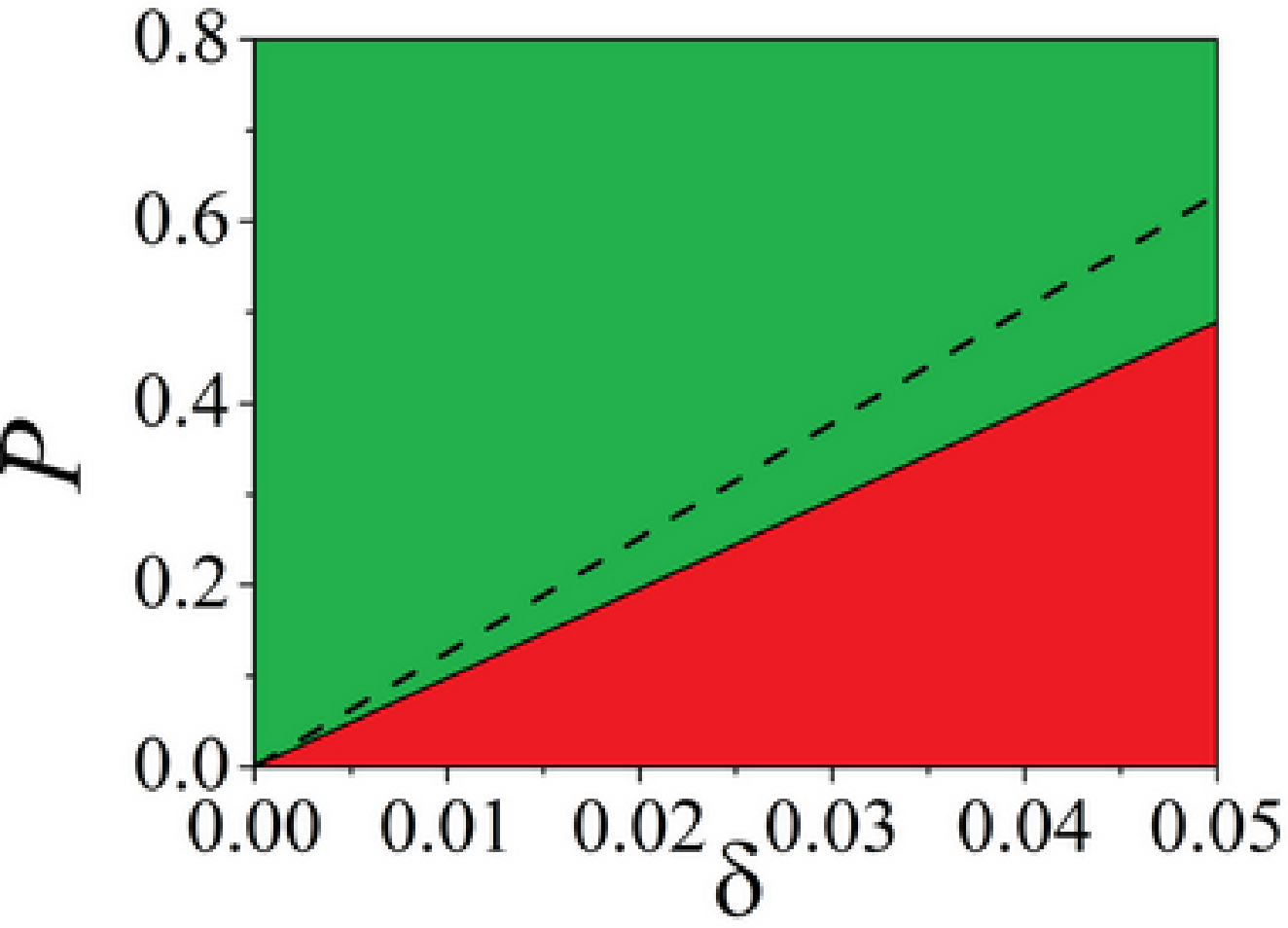}}
\caption{(Color online) The comparison between the theoretically predicted
(dashed lines) and numerically found boundaries of the existence of the
stable 2H-breaking (a,d) and symmetry-breaking (b,c) modes in the
plane of $(\protect\delta $,$P)$, in the interval of $0\leq |\protect\delta %
|\equiv \left\vert 1/2-\protect\omega \right\vert \leq 0.05$.
In the red areas, only even or 2H modes are produced by the numerical
solution, while in the yellow (bistability) regions they coexist with the
stable symmetry- or 2H-breaking modes, respectively. Panels (a)
and (b) pertain, severally, to the SF and SDF nonlinear potentials, while
(c) and (d) correspond to the alternating SF-SDF potential. These four
panels are, actually, zoomed versions of the right bottom corners of Figs.
\protect\ref{fig_4_b}, \protect\ref{fig_7_a}, \protect\ref{fig_8_a}, and
\protect\ref{fig_8_b}, respectively, with analytically predicted lines (%
\protect\ref{SFborder}) or (\protect\ref{SFSDFborder}) added to each panel.}
\label{Comparison}
\end{figure}

The corresponding relation between $\mu $ and $P$ is
\begin{equation}
\mu ={\frac{\delta }{2}+\frac{P}{8\pi }},  \label{mu1}
\end{equation}%
which satisfies the above-mentioned anti-VK criterion, cf. Eq. (\ref{anti}),
i.e., it is possible to assume that the stability of the symmetry-breaking
mode is supported by the SDF part of the alternating nonlinear potential.
Further, Eq. (\ref{SFSDFac}) predicts the emergence of the symmetry-breaking
mode ($a_{0}^{2}=0$) at $\mu =\delta $. In terms of the total power related
to $\mu $ by Eq. (\ref{P}), this implies that this mode exists at
\begin{equation}
P\geq P_{\mathrm{\min }}=4\pi |\delta |,  \label{SFSDFborder}
\end{equation}%
cf. Eq. (\ref{SFborder}). The comparison of this analytical result with its
numerical counterpart in shown in Fig. \ref{Comparison}(c).

The numerical results presented in the previous section demonstrate that the
alternating SF-SDF nonlinear potential support the stable 2H-breaking mode
too. To describe it in the framework of the two-mode approximation, we now
assume $a_{1}$ to be imaginary, as in Eq. (\ref{i}). In this case, Eq. (\ref%
{SFSDFac}) yields
\begin{eqnarray}
&&a_{0}^{2}=2(\delta -\mu ),  \notag \\
&&c^{2}=-2\mu ,  \label{SFSDFac2}
\end{eqnarray}%
cf. Eq. (\ref{SFSDFa0a1}), the respective relation between $\mu $ and $P$
being
\begin{equation}
\mu ={\frac{\delta }{2}}-{\frac{P}{8\pi }},  \label{mu2}
\end{equation}%
cf. Eq. (\ref{mu1}). The latter relation satisfies the VK criterion, which
implies that the stability of the 2H-breaking mode is supported by the SF
part of the alternating nonlinear potential. The emergence of the
2H-breaking mode corresponds to $a_{0}^{2}=0$, i.e., again $\mu =\delta $,
as in the case of the symmetry-breaking mode, under the same alternating
nonlinear potential. Finally, this means that the existence of the stable
2H-breaking mode is predicted in the same region (\ref{SFSDFborder}) as for
its symmetry-breaking counterpart. The latter prediction is compared to the
numerical findings in Fig. \ref{Comparison}(d).

\section{CONCLUSION}

This work aimed to study the existence and stability of modes trapped in the
rotating nonlinear-lattice potentials of the SF, SDF (self-focusing and
defocusing) and alternating SF-SDF types. The consideration was carried out
for the first rotational Brillouin zone in the rotating reference frame, i.e., for $%
0\leq \omega \leq 1/2$, where $\omega $\ is the rotation speed. The
stability analysis was performed through the computation of eigenvalues for
small perturbations, and verified by direct simulations. The model can be
realized in the spatial domain, in terms of a twisted-pipe optical
waveguide, with the built-in azimuthal modulation of the local Kerr
coefficient, or, in the temporal domain, as the Gross-Pitaevskii equation
for BEC loaded into a toroidal trap, under the action of a rotating optical
or magnetic structure which affects the local value of the scattering length.

In the SF system, four types of different modes have been identified: even,
odd, which are dominated by combinations of the fundamental and zeroth
angular harmonic, and the 2H (second-harmonic) and 2H-breaking states. On
the other hand, the SDF nonlinear potential supports three species of the
trapped states: 2H, even, and symmetry-breaking ones, the latter existing
only at the rotation speed close to the right edge of the Brillouin zone, $%
\omega =1/2$, and in a limited interval of values of the total power, $P$.
As for the alternating SF-SDF\ nonlinear potential, it supports all the five
species of the trapped modes. Transitions between the 2H and 2H-breaking
modes are of the supercritical types. The energy comparison reveals that, in
the SF and SF-SDF systems alike, the 2H-breaking mode, if it exists,
represents the ground state; otherwise, this role is played by the 2H mode.
The ground state of the SDF system is always represented by the 2H solution.

This work may be naturally developed in other directions. On the one hand,
it is relevant to consider rotating nonlinear lattices with smaller
azimuthal periods, $2\pi /n$, for integer $n>1$, unlike the case of $n=1$
investigated here. On the other hand, it may be interesting to consider a
two-dimensional version of the present one-dimensional model (with an entire
rotating plane, rather than the thin ring, which will include effects of the
Coriliolis' force).

\begin{acknowledgments}
We appreciate a valuable discussion with Dimitri J. Frantzeskakis, and help
in the use of numerical methods provided by Nir Dror and Shenhe Fu. This
work was supported by CNNSF(grant No. 11104083,11204089,11205063), by the German-Israel Foundation
through grant No. I-1024-2.7/2009, and by the Tel Aviv University in the
framework of the ``matching" scheme.
\end{acknowledgments}

\bibliographystyle{plain}
\bibliography{apssamp}

\begin{thebibliography}{99}
\bibitem{YVKRevMod} Y. V. Kartashov, B. A. Malomed, and L. Torner, Rev. Mod.
Phys. \textbf{83}, 247 (2011).

\bibitem{Harrison} W. A. Harrison, \textit{Pseudopotentials in the Theory of
Metals} (Benjamin: New York, 1966).

\bibitem{LCQian} F. K. Abdullaev, A. Gammal, and L. Tomio, J. Phys. B
\textbf{37}, 635 (2004); H. Sakaguchi and B. A. Malomed, Phys. Rev. E
\textbf{72}, 046610 (2005); F. K. Abdullaev and J. Garnier, Phys. Rev. A
\textbf{72}, 061605(R) (2005); G. Theocharis, P. Schmelcher, P. G.
Kevrekidis, and D. J. Frantzeskakis, \textit{ibid}. \textbf{72}, 033614
(2005); D. L. Machacek, E. A. Foreman, Q. E. Hoq, P. G. Kevrekidis, A.
Saxena, D. J. Frantzeskakis, and A. R. Bishop, Phys. Rev. E \textbf{74},
036602 (2006); P. Niarchou, G. Theocharis, P. G. Kevrekidis, P. Schmelcher,
and D. J. Frantzeskakis, Phys. Rev. A 76, 023615 (2007); Z. Rapti, P. G.
Kevrekidis, V. V. Konotop, and C. K. R. T. Jones, J. Phys. A \textbf{40},
14151 (2007); F. Abdullaev, A. Abdumalikov, and R. Galimzyanov, Phys. Lett.
A \textbf{367}, 149 (2007); F. K. Abdullaev, A. Gammal, M. Salerno, and L.
Tomio, Phys. Rev. A \textbf{77}, 023615 (2008); L. C. Qian, M. L. Wall, S.
Zhang, Z. Zhou, and H. Pu, \textit{ibid}. \textbf{77}, 013611 (2008); A. S.
Rodrigues, P. G. Kevrekidis, M. A. Porter, D. J. Frantzeskakis, P.
Schmelcher, and A. R. Bishop, \textit{ibid}. \textbf{78}, 013611 (2008); F.
K. Abdullaev, Y. V. Bludov, S. V. Dmitriev, P. G. Kevrekidis, and V. V.
Konotop, Phys. Rev. E \textbf{77}, 016604 (2008); Y. V. Kartashov, B. A.
Malomed, V. A. Vysloukh, and L. Torner, Opt. Lett. \textbf{34}, 3625 (2009);
F. K. Abdullaev, R. M. Galimzyanov, M. Brtka, and L. Tomio, Phys. Rev. E
\textbf{79}, 056220 (2009); N. Dror and B. A. Malomed, Phys. Rev. A \textbf{%
83}, 033828 (2011); X.-F. Zhou, S.-L. Zhang, Z.-W. Zhou, B. A. Malomed, and
H. Pu, \textit{ibid}. A \textbf{85}, 023603 (2012).

\bibitem{2D} H. Sakaguchi and B.A. Malomed, Phys. Rev. E \textbf{73}, 026601
(2006); G. Fibich, Y. Sivan, and M. I. Weinstein, Physica D \textbf{217}, 31
(2006); Y. Sivan, G. Fibich, and M. I. Weinstein, Phys. Rev. Lett. \textbf{97%
}, 193902 (2006); Y. V. Kartashov, B. A. Malomed, V. A. Vysloukh, and L.
Torner Opt. Lett. \textbf{34}, 770 (2009); Phys. Rev. A \textbf{80}, 053816
(2009); O. V. Borovkova, B. A. Malomed, Y. V. Kartashov, EPL \textbf{92},
64001 (2010); N. V. Hung, P. Zi\'{n}, M. Trippenbach, and B. A. Malomed,
Phys. Rev. E \textbf{82}, 046602 (2010); T. Mayteevarunyoo, B. A. Malomed,
and A. Reoksabutr, J. Mod. Opt. \textbf{58}, 1977 (2011); H. Sakaguchi and
B. A. Malomed, Opt. Lett. \textbf{37}, 1035 (2012).

\bibitem{singular} T. Mayteevarunyoo, B. A. Malomed, and G. Dong, Phys. Rev.
A \textbf{78}, 053601 (2008); O. V. Borovkova, V. E. Lobanov, and B. A.
Malomed, \textit{ibid}. A \textbf{85}, 023845 (2012); A. Acus, B. A.
Malomed, and Y. Shnir, Physica D \textbf{241}, 987 (2012).

\bibitem{Barcelona} O. V. Borovkova, Y. V. Kartashov, B. A. Malomed, and L.
Torner, Opt. Lett. \textbf{36}, 3088 (2011); O. V. Borovkova, Y. V.
Kartashov, L. Torner, and B. A. Malomed, Phys. Rev. E \textbf{84}, 035602
(R) (2011); W.-P. Zhong, M. Beli\'{c}, G. Assanto, B. A Malomed, and T.
Huang, Phys. Rev. A \textbf{84}, 043801 (2011); Y. V. Kartashov, V. A.
Vysloukh, L. Torner, and B. A. Malomed, Opt. Lett. \textbf{36}, 4587 (2011);
V. E. Lobanov, O. V. Borovkova, Y. V. Kartashov, B. A. Malomed, and L.
Torner, \textit{ibid}. \textbf{37}, 1799 (2012).

\bibitem{Fluan} F. Luan, A. K. George, T. D. Hedley, G. J. Pearce, D. M.
Bird, J. C. Knight, and P. St. J. Russell, Opt. Lett. \textbf{29}, 2369
(2004); G. Bouwmans, L. Bigot, Y. Quiquempois, F. Lopez, L. Provino, and M.
Douay, Opt. Exp. \textbf{13}, 8452 (2005); A. Fuerbach, P. Steinvurzel, J.
A. Bolger, A. Nulsen, and B. J. Eggleton, Opt. Lett. \textbf{30}, 830 (2005).

\bibitem{TTLarsen} T. T. Larsen, A. Bjarklev, D. S. Hermann, and J. Broeng,
Opt. Exp. \textbf{11}, 2589 (2003); F. Du, Y. Q. Lu, and S. T. Wu, Appl.
Phys. Lett. \textbf{85}, 2181 (2004); M. W. Haakestad, T. T. Alkeskjold, M.
D. Nielsen, L. Scolari, J. Riishede, H. E. Engan, and A. Bjarklev, IEEE
Phot. Tech. Lett. \textbf{17}, 819 (2005); C. R. Rosberg, F. H. Bennet, D.
N. Neshev, P. D. Rasmussen, O. Bang, W. Kr\'{o}likowski, A. Bjarklev, and Y.
S. Kivshar, Opt. Exp. \textbf{15}, 12145 (2007).

\bibitem{Kip} J. Hukriede, D. Runde, and D. Kip, J. Phys. D \textbf{36}, R1
(2003).

\bibitem{Liangbing} J. Li, B. Liang, Y. Liu, P. Zhang, J. Zhou, S. O.
Klimonsky, A. S. Slesarev, Y. D. Tretyakov, L. O. Faolain, and T. F. Krauss,
Adv. Mater. \textbf{22}, 1 (2010); M. Feng, Y. Liu, Y. Li, X. Xie, and J.
Zhou, Opt. Exp. \textbf{19}, 7222 (2011); Y. Li, B. A. Malomed, J. Wu, W.
Pang, S. Wang, and J. Zhou, Phys. Rev. A \textbf{84}, 043839 (2011);B.
Liang, Y. Liu, J. Li, L. Song, Y. Li, J. Zhou and K. S. Wong, J. Micromech.
Microeng. \textbf{22} 035013 (2012).

\bibitem{Yongyao} Y. Li, B. A. Malomed, M. Feng, and J. Zhou, Phys. Rev. A
\textbf{82} 063813 (2010); W. Pang, J. Wu, Z. Yuan, Y. Liu and G. Chen, J.
Phys. Soc. Jpn. \textbf{80}, 113401 (2011); J. Wu, M. Feng, W. Pang, S. Fu,
and Y. Li, J. Nonlin. Opt. Phys. \textbf{20}, 193 (2011); Y. Li, W. Pang, S.
Fu, and B. A. Malomed, Phys. Rev. A \textbf{85}, 053821 (2012).

\bibitem{Fleischhauer} M. Fleischhauer, A. Imamo\v{g}lu, and J. P. Marangos,
Rev. Mod. Phys. \textbf{77}, 633 (2005); H. Schmidt and A. Imamo\v{g}lu,
Opt. Lett. \textbf{21}, 1936 (1996); H. Wang, D. Goorskey, and M. Xiao,
Phys. Rev. Lett. \textbf{87}, 073601 (2001).

\bibitem{SInouye} S. Inouye, M. R. Andrews, J. Stenger, H. J. Miesner, D. M.
Stamper-Kurn, and W. Ketterle, Nature (London) \textbf{392}, 151 (1998); Ph.
Courteille, R. S. Freeland, D. J. Heinzen, F. A. van Abeelen, and B. J.
Verhaar, Phys. Rev. Lett. \textbf{81}, 69 (1998); J. L. Roberts, N. R.
Claussen, J. P. Burke, C. H. Greene, E. A. Cornell, and C. E. Wieman,\textit{%
\ ibid}. \textbf{81}, 5109 (1998).

\bibitem{optical} P. O. Fedichev, Y. Kagan, G. V. Shlyapnikov, and J. T. M.
Walraven, Phys. Rev. Lett. \textbf{77}, 2913 (1996); M. Theis, G.
Thalhammer, K. Winkler, M. Hellwig, G. Ruff, R. Grimm, and J. H. Denschlag,
\textit{ibid}. \textbf{93}, 123001 (2004).

\bibitem{magnetic} S. Ghanbari, S., T. D. Kieu, A. Sidorov, and P.
Hannaford, J. Phys. B \textbf{39}, 847 (2006); A. Abdelrahman, P. Hannaford,
and K. Alameh, Opt. Express \textbf{17}, 24358 (2009).

\bibitem{vort-latt} A. E. Fetter, Rev. Mod. Phys. \textbf{81}, 647 (2009).

\bibitem{Cornell} P. Engels, I. Coddington, P. C. Haljan, V. Schweikhard,
and E. A. Cornell, Phys. Rev. Lett. \textbf{90}, 170405 (2003).

\bibitem{giant} A. L. Fetter, Phys. Rev. A \textbf{64}, 063608 (2001); E.
Lundh, \textit{ibid}. \textbf{65}, 043604 (2002); K. Kasamatsu, M. Tsubota,
and M. Ueda, \textit{ibid}. \textbf{66}, 053606 (2002); G. M. Kavoulakis and
G. Baym, New J. Phys. \textbf{5}, 51 (2003); A. Aftalion and I. Danaila,
Phys. Rev. A \textbf{69}, 033608 (2004); U. R. Fischer and G. Baym, Phys.
Rev. Lett. \textbf{90}, 140402 (2003); A. D. Jackson and G. M. Kavoulakis,
Phys. Rev. A \textbf{70}, 023601 (2004); T. K. Ghosh, \textit{ibid}. \textbf{%
69}, 043606 (2004).

\bibitem{street} K. Kasamatsu and M. Tsubota, Phys. Rev. A \textbf{76},
023606 (2009).

\bibitem{rotating-trap} N. K. Wilkin, J. M. F. Gunn, and R. A. Smith, Phys.
Rev. Lett. \textbf{80}, 2265 (1998); B. Mottelson, \textit{ibid}. \textbf{83}%
, 2695 (1999); C. J. Pethick and L. P. Pitaevskii, Phys. Rev. A \textbf{62},
033609 (2000); E. Lundh, A. Collin, and K.-A. Suominen, Phys. Rev. Lett.
\textbf{92}, 070401 (2004); G. M. Kavoulakis, A. D. Jackson, and G. Baym,
Phys. Rev. A \textbf{70}, 043603 (2004); A. Collin, \textit{ibid}. \textbf{73%
}, 013611 (2006); \ A. Collin, E. Lundh, and K.-A. Suominen, Phys. Rev. A
\textbf{71}, 023613 (2005); S. Bargi, G. M. Kavoulakis, and S. M. Reimann,
\textit{ibid}. \textbf{73}, 033613 (2006); H. Sakaguchi and B. A. Malomed,
\textit{ibid}. \textbf{78}, 063606 (2008).

\bibitem{sieve} V. Schweikhard, I. Coddington, P. Engels, S. Tung, and E. A.
Cornell, Phys. Rev. Lett. \textbf{93}, 210403 (2004); S. Tung, V.
Schweikhard, and E. A. Cornell, \textit{ibid}. 97, 240402 (2006).

\bibitem{in-sieve} J. W. Reijnders and R. A. Duine, Phys. Rev. Lett. \textbf{%
93}, 060401 (2004); Phys. Rev. A \textbf{71}, 063607 (2005); H. Pu, L. O.
Baksmaty, S. Yi, and N. P. Bigelow, Phys. Rev. Lett. \textbf{94}, 190401
(2005); H. Pu, L. O. Baksmaty, S. Yi, and N. P. Bigelow, \textit{ibid}.
\textbf{94}, 190401 (2005); R. Bhat, L. D. Carr, and M. J. Holland, \textit{%
ibid}. \textbf{96}, 060405 (2006); K. Kasamatsu and M. Tsubota, \textit{ibid}%
. \textbf{97}, 240404 (2006); M. P. Mink, C. Morais Smith, and R. A. Duine,
Phys. Rev. A \textbf{79}, 013605 (2009).

\bibitem{HS} H. Sakaguchi and B. A. Malomed, Phys. Rev. A \textbf{75},
013609 (2007); \textbf{79}, 043606 (2009).

\bibitem{nucleation} R. A. Williams, S. Al-Assam, and C. J. Foot, Phys. Rev.
Lett. \textbf{104}, 050404 (2010).

\bibitem{twistedPC} G. Kakarantzas, A. Ortigosa-Blanch, T. A. Birks, P. St.
J. Russell, L. Farr, F. Couny, and B. J. Mangan, Opt. Lett. \textbf{28}, 158 (2003).

\bibitem{Russell} G. K. L. Wong, M. S. Kang, H. W. Lee, F. Biancalana, C.
Conti, T. Weiss, and P. St. J. Russell, Science \textbf{337}, 446 (2012).

\bibitem{RYChiao} R. Y. Chiao and Y. S. Wu, Phys. Rev. Lett. \textbf{57},
933 (1986); A. Tomita and R. Y. Chiao, ibid. \textbf{57}, 937 (1986); V. I.
Kopp and A. Z. Genack, Opt. Lett. \textbf{28}, 1876 (2003); M. Ornigotti, G.
Della Valle, D. Gatti, and S. Longhi, Phys. Rev. A \textbf{76}, 023833
(2007).

\bibitem{Sujia} S. Jia and J. W. Fleischer, Phys. Rev. A \textbf{79},
041804(R) (2009).

\bibitem{rocking} B. A. Malomed, Phys. Rev. A \textbf{43}, 410 (1991).

\bibitem{Ueda} H. Saito and M. Ueda, Phys. Rev. Lett. \textbf{93}, 220402
(2004); S. Schwartz, M. Cozzini, C. Menotti, I Carusotto, P. Bouyer, and S.
Stringari, New J. Phys. \textbf{8}, 162 (2006).

\bibitem{guiding} Y. V. Kartashov, B. A. Malomed, and L. Torner, Phys. Rev.
A \textbf{75}, 061602(R) (2007).

\bibitem{WenLuo} L. Wen, H. Xiong, and B. Wu, Phys. Rev. A \textbf{82},
053627 (2010).

%
% NKWilkin : N. K. Wilkin and J. M. F. Gunn, Phys. Rev. Lett. 84, 6 (2000).N. R. Cooper, N.K. Wilkin, and J. M. F. Gunn, ibid. 87, 120405 (2001).

\bibitem{SSB} G. J. Milburn, J. Corney, E. M. Wright, and D. F. Walls, Phys.
Rev. A \textbf{55}, 4318 (1997); I. Zapata, F. Sols, and A. J. Leggett,
\textit{ibid}. \textbf{57}, R28 (1998); E. A. Ostrovskaya, Y. S. Kivshar, M.
Lisak, B. Hall, F. Cattani, and D. Anderson, \textit{ibid}. \textbf{61},
031601(2000); R. D'Agosta, B. A. Malomed, C. Presilla, Phys. Lett. A \textbf{%
275}, 424 (2000); R. K. Jackson and M. I. Weinstein, J. Stat. Phys. \textbf{%
116}, 881 (2004); K. W. Mahmud, H. Perry, and W. P. Reinhardt, Phys. Rev. A.
\textbf{71}, 023615 (2005); H. T. Ng and P. T. Leung, \textit{ibid}. \textbf{%
71}, 013601 (2005); T. Schumm, S. Hofferberth, L. M. Andersson, S.
Wildermuth, S. Groth, I. Bar-Joseph, J. Schmiedmayer and P. Kr\"{u}ger,
Nature Physics \textbf{1}, 57 (2005); M. Albiez, R. Gati, J. F\"{o}lling, S.
Hunsmann, M. Cristiani, and M. K. Oberthaler, Phys. Rev. Lett. \textbf{95},
010402 (2005).

\bibitem{we} Y. Li, W. Pang, and B. A. Malomed, Phys. Rev. A {\bf 86} 023832 (2012).

\bibitem{torus} S. Gupta, K. W. Murch, K. L. Moore, T. P. Purdy, and D. M.
Stamper-Kurn, Phys. Rev. Lett. \textbf{95}, 143201 (2005); A. S. Arnold, C.
S. Garvie, and E. Riis, Phys. Rev. A 73, 041606(R) (2006); I. Lesanovsky and
W. von Klitzing, \textit{ibid}. \textbf{99}, 083001 (2007); C. Ryu, M. F.
Andersen, P. Clad\'{e}, V. Natarajan, K. Helmerson,and W. D. Phillips, Phys.
Rev. Lett. \textbf{99}, 260401 (2007); A. Ramanathan, K. C. Wright, S. R.
Muniz, M. Zelan, W. T. Hill III, C. J. Lobb, K. Helmerson, W. D. Phillips,
and G. K. Campbell, \textit{ibid}. \textbf{106}, 130401 (20011).

\bibitem{Salasnich} L. Salasnich, A. Parola, and L. Reatto, Phys. Rev. A
\textbf{74}, 031603(R) (2006); R. Kanamoto, H. Saito, and M. Ueda, \textit{%
ibid}. \textbf{73}, 033611 (2006); M. Modugno, C. Tozzo, and F. Dalfovo,
\textit{ibid}. \textbf{74}, 061601 (R)\ (2006); A. V. Carpentier and H.
Michinel, EPL \textbf{78}, 10002 (2007); P. Mason and N. G. Berloff, Phys.
Rev. A \textbf{79}, 043620 (2009); J. Brand, T. J. Haigh, and U. Z\"{u}%
licke, \textit{ibid}. \textbf{80}, 011602 (R) (2009); P. Capuzzi and D. M.
Jezek, J. Phys. B: At. Mol. Opt. Phys. \textbf{42}, 145301 (2009); A.
Aftalion and P. Maso, \textit{ibid}. \textbf{81}, 023607 (2010); J.
Smyrnakis, M. Magiropoulos, G. M. Kavoulakis, and A. D. Jackson, \textit{ibid%
}. \textbf{81}, 063601 (2010); S. Z\"{o}llner, G. M. Bruun, C. J. Pethick,
and S. M. Reimann, Phys. Rev. Lett. \textbf{107}, 035301 (2011); Z.-W. Zhou,
S.-L. Zhang, X.-F. Zhou, G.-C. Guo, X. Zhou, and H. Pu, Phys. Rev. A \textbf{%
83}, 043626 (2011); M. Abad, M. Guilleumas, R. Mayol, M. Pi, and D. M.
Jezek, \ \textit{ibid}. \textbf{84}, 035601 (2011); X. Zhou, S. Zhang, Z.
Zhou, B. A. Malomed, and H. Pu, \textit{ibid}. \textbf{85}, 023603 (2012);
S. K. Adhikari, \textit{ibid}. \textbf{85}, 053631 (2012).

\bibitem{Stringari} S. Schwartz, M. Cozzini, C. Menotti, I Carusotto, P.
Bouyer, and S. Stringari, New J. Phys. \textbf{8}, 162 (2006); J. Smyrnakis,
S. Bargi, G. M. Kavoulakis, M. Magiropoulos, K. K\"{a}rkk\"{a}inen, and S.
M. Reimann, Phys. Rev. Lett. \textbf{103}, 100404 (2009).

\bibitem{YJK} J. Yang and T. I. Lakoba, Stud. Appl. Math. \textbf{118}, 153
(2007); \textbf{120}, 265 (2008).

\bibitem{bif} G. Iooss and D. D. Joseph, \textit{Elementary Stability and
Bifurcation Theory} (Springer, New York: 1980).

\bibitem{VK} M. Vakhitov and A. Kolokolov, Izvestiya VUZov Radiofizika
\textbf{16}, 1020 (1973) [in Russian; English translation: Radiophys.
Quantum. Electron. \textbf{16}, 783 (1973)]; L. Berg\'{e}, Phys. Rep.
\textbf{303}, 259 (1998); E. A. Kuznetsov and F. Dias, \textit{ibid}.
\textbf{507}, 43 (2011).

\bibitem{anti} H. Sakaguchi and B. A. Malomed, Phys. Rev. A \textbf{81},
013624 (2010).

%\bibitem{BSHam}B. S. Ham, P. R. Hemmer, M. K. Kim, and S. M. Shahriar: Laser Phys.
%{\bf 9} 788 (1999); A. V. Turukhin, V. S. Sudarshanam, M. S. Shahriar, J. A. Musser,
%B. S. Ham, and P. R. Hemmer: Phys. Rev. Lett. {\bf 88} 023602 (2002); J. J. Longdell, E. Fraval, M. J. Sellars, and N. B. Manson, ibid. {\bf 95} 063601(2005).
\end{thebibliography}
% Produces the bibliography via BibTeX.

\end{document}